\documentclass[12pt,a4paper]{article}
\usepackage[utf8]{inputenc}
\usepackage{epsf}
\usepackage{latexsym,amssymb,euscript}
\usepackage[dvips]{graphicx}
\usepackage[numbers,sort&compress]{natbib}
\usepackage{amsmath}
\usepackage{nicefrac}
\usepackage{slashed}
\usepackage{booktabs}
\usepackage[linktocpage]{hyperref}
\usepackage{braket}
\usepackage{chngcntr}
\usepackage{bbold}
\usepackage{multicol}
\usepackage{graphics}
\usepackage{graphicx}
\usepackage{mciteplus}
\usepackage{caption}
\usepackage{subcaption}
\usepackage{pdfpages}
\usepackage[titletoc]{appendix}
\usepackage[normalem]{ulem}
\graphicspath{{./figures/}}
\hypersetup{
 linktocpage = false,
 urlcolor = urlblue,
 colorlinks = true,
 linkcolor = urlblue,
 anchorcolor = linkred,
 citecolor = citegreen,
 pdfstartview = {XYZ null null 1.25} 
           }
\usepackage[left=2cm, right=2cm]{geometry}
\usepackage{pstricks}
\usepackage{color}
\usepackage{xcolor}
\definecolor{urlblue}{rgb}{0.2,0.4,0.7}
\definecolor{citegreen}{rgb}{0,0.4,0.2}
\definecolor{linkred}{rgb}{0.9,0.2,0.1}
\usepackage{float}
\usepackage{academicons}
\definecolor{orcidlogocol}{HTML}{A6CE39}
\usepackage{changepage}
\usepackage{fancyhdr}
\newcommand{\drv}{{\rm d}}

\newcommand{\LQCD}{\Lambda_{\rm QCD}}
\newcommand{\MSb}{\overline{\rm MS}}

\newcommand{\LL}{{\rm LL/LO}}

\newcommand{\NLLp}{{\rm NLL/NLO^+}}
\newcommand{\NLLpp}{{\rm NLL/NLO^{(+)}}}
\newcommand{\HENLOp}{{\rm HE}\mbox{-}{\rm NLO^+}}

\newcommand{\CmLL}{{\cal C}_m^\LL}

\newcommand{\CmNLLp}{{\cal C}_m^\NLLp}

\newcommand{\CmHENLOp}{{\cal C}_m^{{\rm HE}\text{-}{\rm NLO}^+}}

\newcommand{\DY}{\Delta Y}

\newcommand{\vqTTa}{\langle {\vec q}_T^{\;2} \rangle}

\newcommand{\E}{{\cal E}}

\newcommand{\Jpsi}{J/\psi}

\newcommand{\BCs}{B_c(^1S_0)}
\newcommand{\Bss}{B_c(^3S_1)}

\newcommand{\XQq}{X_{Qq\bar{Q}\bar{q}}}

\newcommand{\TQQ}{T_{4Q}}
\newcommand{\TQc}{T_{4c}}

\newcommand{\PQQ}{P_{5Q}}
\newcommand{\PQc}{P_{5c}}

\newcommand{\bPQQ}{\bar{P}_{5Q}}
\newcommand{\bPQc}{\bar{P}_{5c}}

\newcommand{{\HFNRevo}}{\tt HF-NRevo}

\newcommand{{\Jethad}}{\tt JETHAD}
\newcommand{{\symJethad}}{\tt symJETHAD}
\newcommand{{\psymJethad}}{\tt (sym)JETHAD}
\newcommand{{\Hell}}{\tt HELL}
\newcommand{{\RadISH}}{\tt RadISH}
\newcommand{{\Pegasus}}{\tt QCD-PEGASUS}
\newcommand{{\HOPPET}}{\tt HOPPET}
\newcommand{{\QCDNUM}}{\tt QCDNUM}
\newcommand{{\APFEL}}{\tt APFEL}
\newcommand{{\APFELpp}}{\tt APFEL++}
\newcommand{{\APFELppp}}{\tt APFEL(++)}
\newcommand{{\EKO}}{\tt EKO}
\newcommand{{\FeynCalc}}{\tt FeynCalc}

\newcommand{\eref}[1]{~\eqref{#1}}

\newcommand{\orcidFGC}{\href{https://orcid.org/0000-0003-3299-2203}{\includegraphics[scale=0.1]{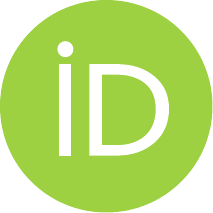}}}

\begin{document}

\begin{titlepage}

{
\begin{adjustwidth}{-1cm}{-1cm}
\begin{center}
  {\Large \bf Multimodal Fragmentation of All-Heavy Pentaquarks: \vskip.035cm Uncertainty-Aware Predictions for Hadron Colliders}
\end{center}
\end{adjustwidth}
}

\vskip 0.75cm

\centerline{
Francesco~Giovanni~Celiberto$^{\;1\;\dagger}$ {\orcidFGC}
}

\vskip .4cm

\centerline{${}^1$ {\sl Universidad de Alcal\'a (UAH), Departamento de Física y Matemáticas,}}
\centerline{\sl E-28805 Alcal\'a de Henares, Madrid, Spain}
\vskip 1.15cm

\begin{abstract}
\vspace{0.25cm}
\headrule 
\vspace{0.50cm}
We present an uncertainty-aware description of leading-power fragmentation for all-charm pentaquark states ($S$-wave $|cccc\bar{c}\rangle$) at hadron colliders. 
We construct a multimodal set of collinear fragmentation functions, {\tt PQ5Q1.1}, incorporating both perturbative and nonperturbative uncertainties. 
Perturbative effects are estimated via missing higher-order variations (F-MHOUs), while the nonperturbative wave function is modeled through controlled modifications of its transverse-momentum structure (F-NPWF), consistently combined within a replicalike framework. 
The initial-scale input for constituent charm fragmentation is refined to describe both compact multiquark and diquark-driven production mechanisms. 
We employ the {\tt (sym)JETHAD} interface to study NLL/NLO$^+$ semi-inclusive pentaquark-plus-jet production at the HL-LHC and future FCC. 
The bottom sector is left to future dedicated studies due to its enhanced sensitivity to nonperturbative modeling. 
Our results provide a flexible framework for uncertainty-controlled predictions, bridging exotic-hadron structure, heavy-flavor fragmentation, and high-energy QCD.
\vspace{0.50cm} 
\headrule
\vspace{0.75cm}
{
 \setlength{\parindent}{0pt}
 \textsc{Keywords}: \vspace{0.15cm} \\ 
 Exotic matter \\
 All-charm pentaquarks \\
 Precision QCD \\
 Multimodal fragmentation \\
 Hadronic structure \\
 Heavy flavor \\
 Resummation \\
 {\tt PQ5Q1.1} FF release
\vspace{0.65cm} 
\headrule
}
\end{abstract}

\vspace{-0.00cm}
\vfill
$^{\dagger}${\it e-mail}:
\href{mailto:francesco.celiberto@uah.es}{francesco.celiberto@uah.es}

\end{titlepage}

\tableofcontents
\clearpage

\section{Introductory aspects}
\label{sec:intro}

The exploration of multiquark hadrons has profoundly broadened our view of hadronic matter, revealing a spectrum far richer than that envisioned in the conventional quark model~\cite{Brambilla:2019esw,Yuan:2018inv,JPAC:2021rxu}. 
While the Gell-Mann--Zweig classification identifies mesons as quark-antiquark states and baryons as three-quark systems~\cite{Gell-Mann:1964ewy,Zweig:1964jf}, a large body of experimental and theoretical evidence now supports the existence of nonconventional configurations, including tetraquarks, pentaquarks, hexaquarks, and hybrid mesons~\cite{Jaffe:1976ih,Jaffe:1976yi,Rosner:1985yh,Pepin:1998ih,Vijande:2011im,Esposito:2016noz,Lebed:2016hpi,Guo:2017jvc,Lucha:2017mof,Ali:2019roi}. 
These states offer a privileged laboratory for investigating the nonperturbative dynamics of Quantum Chromodynamics (QCD).

Future high-energy facilities, such as the High-Luminosity Large Hadron Collider (HL-LHC)~\cite{Apollinari:2015wtw}, the Electron-Ion Collider (EIC)~\cite{AbdulKhalek:2021gbh,Khalek:2022bzd,Abir:2023fpo}, and the Future Circular Collider (FCC)~\cite{FCC:2018byv,FCC:2018evy,FCC:2018vvp,FCC:2018bvk}, will considerably extend the kinematic reach of exotic-hadron searches. 
In parallel, the combination of fixed-order calculations with all-order resummation techniques has enabled increasingly precise predictions for heavy-hadron production, allowing for a consistent treatment of multiscale dynamics and for systematic comparisons with data. 
Within this broader program, exotic spectroscopy and precision QCD can be naturally merged into a unified framework for studying color confinement and multiquark bound states.

Several scenarios have been proposed for the internal structure of exotic hadrons. 
In the compact multiquark picture, they are interpreted as tightly bound quark-antiquark aggregates~\cite{Jaffe:2003sg,Maiani:2015vwa}. 
Alternative descriptions include the hadronic molecular model, where states emerge as loosely bound meson-meson or meson-baryon systems near threshold~\cite{Tornqvist:1993ng,Braaten:2003he,Guo:2013sya,Mutuk:2022ckn,Wang:2014gwa,Guo:2017jvc,Karliner:2015ina,Chen:2016qju,Esposito:2023mxw,Grinstein:2024rcu}, the diquark-antidiquark framework~\cite{Maiani:2004vq,tHooft:2008rus,Maiani:2013nmn,Maiani:2014aja,Maiani:2017kyi,Wang:2013exa,Grinstein:2024rcu}, and the hadroquarkonium model, where a heavy quarkonium core is embedded in a light-quark environment~\cite{Dubynskiy:2008mq,Voloshin:2013dpa,Eides:2019tgv,Ferretti:2018tco,Ferretti:2020ewe}. 
The latter has been successfully applied to states such as $X(3872)$~\cite{Choi:2003ue} and $Y(4260)$~\cite{CLEO:2006ike}.

A major step forward in the study of all-heavy systems has been the observation of fully charmed tetraquark candidates, $X(6600)$, $X(6900)$, and $X(7100)$, in double-$\Jpsi$ and $\Jpsi$-plus-$\psi(2S)$ channels~\cite{CMS:2023owd}. 
Located in the region $4m_c < M < 7\,{\rm GeV}$, these structures currently represent the only established fully heavy exotics. 
Their interpretation has been reinforced by recent spin-parity analyses~\cite{CMS:2025fpt}, which favor a $J^{PC} = 2^{++}$ assignment for the leading structure and disfavor molecular explanations in favor of compact multiquark configurations. 
This provides strong motivation to extend the investigation to other all-heavy states.

The experimental search for pentaquarks has followed a more intricate path. 
Because the antiquark entering the leading Fock state can cancel in flavor with one of the quarks, some pentaquark signatures may mimic ordinary baryons, making their identification particularly challenging. 
For this reason, early searches concentrated on channels where such cancellations do not occur. 
The first claims date back to 2003, when the LEPS and DIANA Collaborations reported evidence for the $\Theta^+$ baryon~\cite{LEPS:2003wug,DIANA:2003uet}, in line with earlier theoretical expectations~\cite{Diakonov:1997mm}. 
However, subsequent null results and conflicting measurements led to substantial skepticism~\cite{CLAS:2005koo,Hicks:2004ge,ParticleDataGroup:2008zun,Praszalowicz:2024mji}.

A decisive turning point was reached in 2015, when the LHCb Collaboration observed two charmonium-like pentaquark candidates, $P_c(4380)^+$ and $P_c(4450)^+$, in $\Lambda_b$ decays~\cite{Aaij:2015tga}. 
These states, reconstructed in the $[J/\psi \,+\, p]$ channel, provided the first unambiguous evidence of pentaquark structures. 
Subsequent analyses led to the discovery of additional candidates, including $P_c(4312)^+$, $P_c(4440)^+$, and $P_c(4457)^+$~\cite{Aaij:2019vzc}, as well as the strange pentaquark candidate $P_{cs}(4459)$~\cite{LHCb:2020jpq} and the $P_{\psi_s}^{\Lambda}(4338)$ resonance~\cite{LHCb:2022ogu}. 
Further experimental developments are discussed in Refs.~\cite{LHCb:2016lve,LHCb:2021chn}.

Despite this progress in spectroscopy, the dynamical mechanisms responsible for pentaquark production remain poorly understood. 
Only a limited number of model-dependent approaches have been explored so far, including color evaporation~\cite{Maciula:2020wri} and hadron-quark-duality descriptions~\cite{Karliner:2016zzc,Becchi:2020mjz}. 
At the same time, lattice QCD and effective field theories have clarified several aspects of interquark forces and resonance formation~\cite{Francis:2018jyb,Leskovec:2019ioa,Liu:2019tjn,Bicudo:2022cqi,Alexandrou:2024iwi,Prelovsek:2023sta}, while more recent studies have emphasized the role of heavy-quark dynamics, meson-baryon interactions~\cite{Xiao:2019aya}, and the interplay between long-range pion exchange and short-distance quark forces~\cite{Ali:2017jda,Yamaguchi:2017zmn}.

From a theoretical perspective, all-heavy multiquark systems provide a particularly clean environment. 
Here, the heavy-quark mass $m_Q$ sets the dominant scale, while the absence of light valence quarks suppresses genuinely soft contributions. 
An all-heavy tetraquark can then be described by the lowest Fock state $|QQ\bar{Q}\bar{Q}\rangle$, closely resembling the quarkonium configuration $|Q\bar{Q}\rangle$. 
This analogy motivates the extension of quarkonium-inspired methods to exotic systems, often viewed as QCD analogs of few-body atomic configurations~\cite{Pineda:2011dg}. 
In the same spirit, an all-charm pentaquark $\PQc$ can be modeled as a nonrelativistic bound state $| c \bar{c} c c c \rangle$.

Measurements of large transverse-momentum production rates, such as those observed for the $X(3872)$~\cite{CMS:2013fpt,ATLAS:2016kwu,LHCb:2021ten}, suggest that fragmentation mechanisms may play an important role in exotic-hadron production. 
This motivates the use of leading-power factorization frameworks, in which a single parton fragments into the observed hadron, thereby providing a direct bridge between high-energy QCD dynamics and hadron formation.

In this direction, a new generation of fragmentation-function (FF) frameworks has been developed to describe heavy and exotic hadrons in a unified way. 
These approaches combine model-based calculations of the initial-scale inputs~\cite{Suzuki:1977km,Nejad:2021mmp,Feng:2020riv,Bai:2024ezn,Bai:2024flh} with DGLAP evolution schemes that consistently account for heavy-quark thresholds~\cite{Mele:1990cw,Cacciari:1993mq,Celiberto:2025euy,Celiberto:2024mex,Celiberto:2024bxu,Celiberto:2024rxa,Celiberto:2025xvy,Celiberto:2026rzi,Celiberto:2026zss}. 
This program has led to several FF families, including {\tt TQHL1.1} and {\tt TQ4Q1.1}({\tt 2.0}) for doubly and fully heavy tetraquarks~\cite{Celiberto:2024beg,Celiberto:2025dfe,Celiberto:2025ziy,Celiberto:2026kks}.
Overall, these developments provide a flexible and systematically improvable basis for studying exotic-hadron production across different kinematic regimes.

In this context, heavy-flavor fragmentation at leading power naturally lies at the interface between hadronic structure and precision QCD. 
The presence of heavy quarks introduces a dual dynamics: short-distance production can be treated perturbatively, while bound-state formation requires modeling of nonperturbative effects.

Building on these developments, a natural next step is to extend this framework to the pentaquark sector.

In this work, we develop a general framework for the leading-power fragmentation of all-heavy pentaquark states and present its first phenomenological application to the all-charm sector ($S$-wave $|c\bar{c}ccc\rangle$ pentacharms, hereafter $\PQc$) at present and future hadron colliders. 
Within this context, we introduce a new generation of collinear fragmentation functions, the {\tt PQ5Q1.1} set, which extends and supersedes the previous {\tt PQ5Q1.0} determination~\cite{Celiberto:2025ipt}. 
The {\tt 1.1} update promotes the \emph{multimodal} structure of the initial-scale input, where distinct production mechanisms---from compact multiquark configurations to diquark-driven structures---are consistently embedded within a unified fragmentation framework. 
This development marks a transition from exploratory modeling toward a first uncertainty-controlled description of pentacharm fragmentation.

A key advancement of the {\tt PQ5Q1.1} framework is its uncertainty-aware construction. 
While {\tt PQ5Q1.0} was restricted to central-value predictions, here we systematically incorporate both perturbative and nonperturbative sources of uncertainty. 
Perturbative effects are assessed through missing higher-order variations (F-MHOUs), while the nonperturbative wave function (F-NPWF) is probed via controlled modifications of its transverse-momentum structure. 
These contributions are consistently combined within a replicalike framework, enabling a coherent propagation of theoretical uncertainties to phenomenological observables.

This strategy builds on and extends recent developments in heavy-hadron fragmentation. 
In the context of fully heavy tetraquarks, the {\tt TQ4Q1.1} framework~\cite{Celiberto:2025ziy} has demonstrated how the combined treatment of perturbative and nonperturbative uncertainties can provide a robust reference for experimental analyses. 
Similarly, for ordinary heavy-flavor systems, the {\tt NRFF1.0} approach~\cite{Celiberto:2025euy} has established a pathway toward precision studies of (pseudoscalar) quarkonium production by consistently incorporating both sources of uncertainty. 
The present work generalizes this philosophy to the pentaquark sector, providing the first uncertainty-controlled construction in this context.

From a physical standpoint, the need for a multimodal and uncertainty-aware description is deeply rooted in the nature of exotic bound states. 
Even when produced as compact configurations, all-heavy systems can dynamically evolve through quantum fluctuations into more extended structures, such as diquark-based or molecularlike arrangements~\cite{Sazdjian:2022kaf}. 
A first-principles description of these transitions would require solving a genuinely complex multibody bound-state problem, which remains beyond current theoretical control. 
In this context, the {\tt PQ5Q1.1} functions provide an effective and flexible framework to encode such dynamical effects at the level of fragmentation inputs.

Overall, this work contributes to establishing a unified framework for the study of exotic-hadron production at high energies, where hadron-structure modeling, precision QCD techniques, and collider phenomenology are consistently intertwined. 
By promoting a systematic treatment of theoretical uncertainties, it opens the way toward precision-oriented investigations of exotic matter, with direct implications for future experimental analyses at the HL-LHC and FCC.

Although the formal construction of the {\tt PQ5Q1.1} framework is general and applicable to all-heavy systems, our phenomenological analysis is restricted here to the all-charm case. 
Exploratory studies in the bottom sector indicate that the current implementation does not yet provide a sufficiently robust and stable description, with features that remain difficult to interpret in physical terms~\cite{Celiberto:2026ooh}. 
For this reason, a dedicated bottom-sector analysis is deferred to future work.

To connect this framework with collider observables, for the phenomenological analysis we adopt the $\NLLp$ hybrid-factorization scheme, where high-energy logarithms are systematically resummed up to NLL accuracy and matched to next-to-leading-order (NLO) collinear dynamics, including higher-order (NLL$^+$) contributions. 
This framework provides a consistent description of semi-inclusive observables, combining energy resummation with standard collinear factorization.

In the following, we employ the {\Jethad} numerical framework, complemented by its symbolic module {\symJethad}~\cite{Celiberto:2020wpk,Celiberto:2022rfj,Celiberto:2023fzz,Celiberto:2024mrq,Celiberto:2024swu,Celiberto:2025csa}, to derive predictions for high-energy observables sensitive to pentacharm-plus-jet production. 
Our analysis covers center-of-mass energies from the 13~TeV HL-LHC up to the 100~TeV FCC regime.

The paper is organized as follows. 
Section~\ref{sec:FFs} details the construction of the {\tt PQ5Q1.1} collinear FFs for $\PQc$ states. 
In Sec.~\ref{sec:hybrid_factorization}, we outline the $\NLLp$ hybrid-factorization framework. 
Section~\ref{sec:phenomenology} presents our phenomenological results for semi-inclusive $\PQc$ plus jet production at high energy. 
Conclusions and future directions are given in Sec.~\ref{sec:conclusions}.

\section{Multimodal fragmentation of all-charm pentaquarks}
\label{sec:FFs}

In this section we outline our strategy to construct the \emph{multimodal} {\tt PQ5Q1.1} FF family, describing the collinear fragmentation of $S$-wave pentacharms from initial-scale inputs for the charm-quark channel based on either the direct or the scalar-diquark picture.

For completeness, in Sec.~\ref{ssec:FFs-intro} we briefly review the main features of heavy-flavor fragmentation, from heavy-light hadrons to quarkonia and exotic states.
We then focus on the pentaquark case, presenting the direct and diquark inputs in Secs.~\ref{ssec:FFs-Q-direct} and~\ref{ssec:FFs-Q-diquark}, respectively.
Finally, the scale dependence of the {\tt PQ5Q1.1} functions, together with the associated uncertainty estimates from F-MHOU and F-NPWF effects, is presented and discussed in Sec.~\ref{ssec:FFs-PQ5Q11}.

All symbolic computations required to build the {\tt PQ5Q1.1} set are performed using {\symJethad}, a \textsc{Mathematica}~\cite{Mathematica_V14-2} plugin of {\Jethad}~\cite{Celiberto:2020wpk,Celiberto:2022rfj,Celiberto:2023fzz,Celiberto:2024mrq,Celiberto:2024swu,Celiberto:2025csa}, designed for the symbolic manipulation of analytic formulas in hadronic structure and precision QCD.

\subsection{An overview of heavy-flavor fragmentation}
\label{ssec:FFs-intro}

At variance with light hadrons, the fragmentation mechanism underlying heavy-flavor hadronization involves a richer structure, since heavy-quark masses in their lowest Fock state lie within the perturbative QCD domain.
As a consequence, while light-hadron FFs are purely nonperturbative, the initial-scale inputs for heavy-hadron FFs can contain a perturbatively calculable component.

For singly heavy hadrons, such as $D$, $B$, and $\Lambda_{c,b}$ states, fragmentation can be described as a two-stage process~\cite{Cacciari:1996wr,Cacciari:1993mq,Helenius:2023wkn,Generet:2023vte}.
A high-$p_T$ parton $i$, produced in the hard scattering, first fragments into a heavy quark $Q$, a step governed by short-distance dynamics and computable in perturbation theory thanks to the smallness of $\alpha_s(m_Q)$.
This contribution, identified as the short-distance coefficient (SDC) for $[i \to Q]$, occurs over a shorter timescale than hadronization.
The first NLO calculations of these SDCs were performed in Refs.~\cite{Mele:1990yq,Mele:1990cw}, with subsequent next-to-NLO extensions in Refs.~\cite{Mitov:2006wy,Blumlein:2006rr,Melnikov:2004bm,Mitov:2004du,Biello:2024zti}.
At longer timescales, the heavy quark undergoes nonperturbative hadronization, typically modeled through phenomenological approaches~\cite{Bowler:1981sb,Peterson:1982ak,Colangelo:1992kh} or effective field theories~\cite{Georgi:1990um,Eichten:1989zv,Neubert:1993mb}.

A complete ZM-VFNS description is achieved by evolving these initial conditions via the DGLAP equations.
Numerical solutions of the coupled evolution system allow one to consistently incorporate energy-scale dependence at the desired perturbative accuracy.

This two-stage paradigm naturally extends to quarkonia, where the presence of both $Q$ and $\bar{Q}$ in the lowest Fock state $|Q\bar{Q}\rangle$ increases the theoretical complexity.
Modern treatments rely on NRQCD~\cite{Caswell:1985ui,Thacker:1990bm,Bodwin:1994jh}, which provides a factorized description in terms of perturbative SDCs and nonperturbative long-distance matrix elements (see Refs.~\cite{Grinstein:1998xb,Kramer:2001hh,Lansberg:2005aw,Lansberg:2019adr}).
Within this framework, quarkonium states are expanded over a complete set of Fock configurations, ordered through a double expansion in $\alpha_s$ and the relative velocity $v_{\cal Q}$.

The interplay between production mechanisms depends strongly on the transverse momentum $p_T$.
At low $p_T$, quarkonia are dominantly formed via short-distance production of the $(Q\bar{Q})$ pair followed by hadronization.
At larger $p_T$, single-parton fragmentation becomes increasingly relevant and eventually dominates.
The former can be interpreted as a fixed-flavor number-scheme higher-power contribution~\cite{Alekhin:2009ni,Kang:2014tta,Boer:2023zit,Celiberto:2025euy,Celiberto:2024mex,Celiberto:2024bxu,Celiberto:2024rxa,Celiberto:2025xvy,Celiberto:2026rzi,Celiberto:2026zss}, while the latter corresponds to a ZM-VFNS mechanism governed by DGLAP evolution.

Initial-scale FFs for gluon and heavy-quark fragmentation into $S$-wave vector quarkonia were computed at leading order (LO) in Refs.~\cite{Braaten:1993rw,Braaten:1993mp} and later extended to NLO in Ref.~\cite{Zheng:2019gnb}.
These results enabled the construction of the first DGLAP-evolved FF sets for quarkonia, {\tt ZCW19$^+$}\cite{Celiberto:2022dyf,Celiberto:2023fzz}, followed by the {\tt ZCFW22} extension to $\BCs$ and $\Bss$ states~\cite{Celiberto:2022keu,Celiberto:2024omj}.
Phenomenological applications of {\tt ZCFW22} showed that the $\BCs$ production rate relative to singly bottomed $B$ mesons remains below 0.1\%, in agreement with LHCb observations~\cite{LHCb:2014iah,Celiberto:2024omj}, thus supporting the validity of the ZM-VFNS fragmentation picture at high $p_T$.

Extending these ideas to exotic hadrons, NRQCD factorization has been employed to describe double $\Jpsi$ structures observed at the LHC~\cite{LHCb:2020bwg,ATLAS:2023bft,CMS:2023owd}, often interpreted as compact all-charm tetraquarks~\cite{Zhang:2020hoh,Zhu:2020xni}.
In this case, $\TQc$ production originates from the short-distance creation of two charm and two anticharm quarks at a scale $\sim 1/m_c$, followed by nonperturbative binding.
The first NRQCD-based initial condition for gluon fragmentation into color-singlet $S$-wave $\TQc$ states was derived in Ref.~\cite{Feng:2020riv}.

Building on this result, the first VFNS FF family for heavy-light tetraquarks, {\tt TQHL1.0}, was introduced in Ref.~\cite{Celiberto:2023rzw} (see Ref.~\cite{Celiberto:2024mrq} for a review).
Subsequent developments led to the {\tt TQ4Q1.1} and {\tt TQHL1.1} sets~\cite{Celiberto:2024beg}, incorporating NRQCD-based inputs for the $[Q \to \TQQ]$ channel~\cite{Bai:2024ezn}, improved treatments of doubly heavy systems, and bottomoniumlike configurations.
These FFs were later employed in phenomenological studies of all-heavy tetraquark production~\cite{Celiberto:2025ziy,Celiberto:2025dfe,Celiberto:2026zed}, indirect searches in electroweak and Higgs decays~\cite{Ma:2025ryo}, and forward-rapidity analyses sensitive to intrinsic charm~\cite{Celiberto:2025vra}, as well as in dedicated studies of the gluon channel~\cite{Nakhaei:2025zty}.

More recently, the {\tt NRFF1.0} framework~\cite{Celiberto:2025euy} has provided a systematic determination of pseudoscalar-quarkonium FFs with consistent NRQCD inputs, while the {\tt TQ4Q2.0} release extended the multimodal approach to all-heavy tetraquarks by incorporating additional production mechanisms and a refined treatment of uncertainties~\cite{Celiberto:2026kks}.
Within the same research program, first FFs for all-charm pentaquarks and rare $\Omega$ baryons were developed in Refs.~\cite{Celiberto:2025ipt,Celiberto:2025ogy,Celiberto:2025csa}, leading to the {\tt PQ5Q1.0} and {\tt OMG3Q1.0} sets.

\subsection{Direct fragmentation mode}
\label{ssec:FFs-Q-direct}

\begin{figure*}[!t]
\centering
\includegraphics[width=0.475\textwidth]{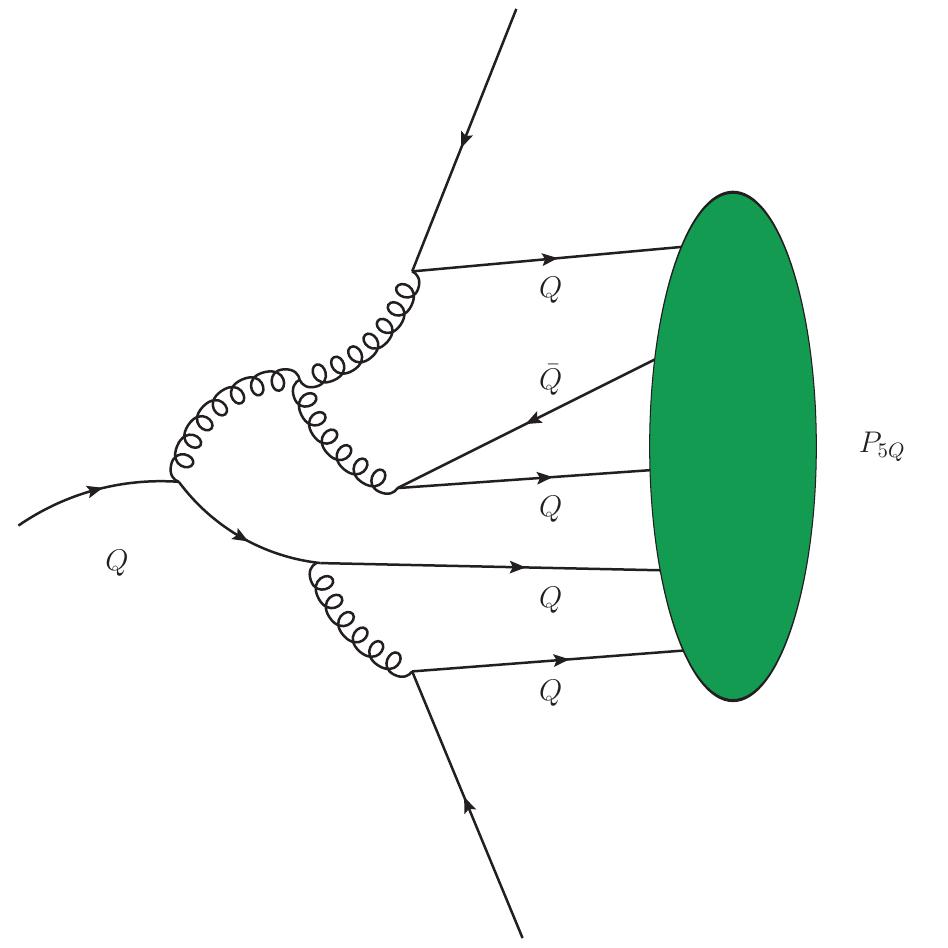}
\hspace{0.40cm}
\includegraphics[width=0.475\textwidth]{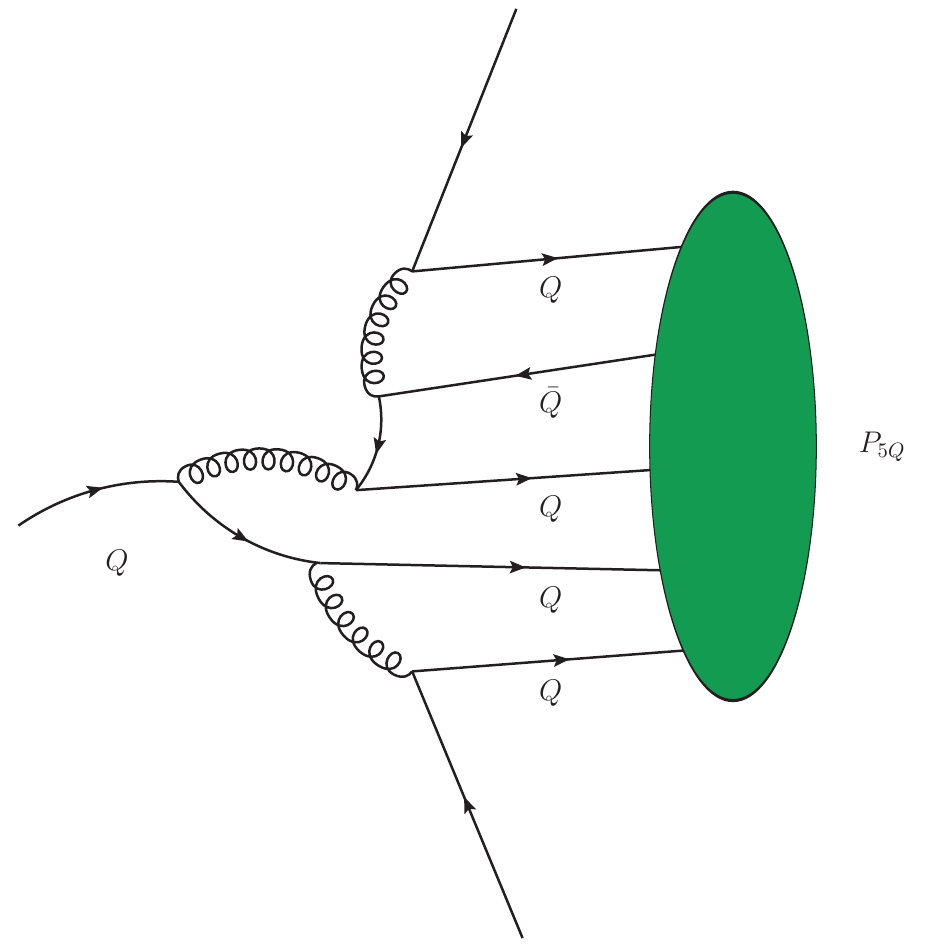}

\vspace{0.20cm}

\includegraphics[width=0.475\textwidth]{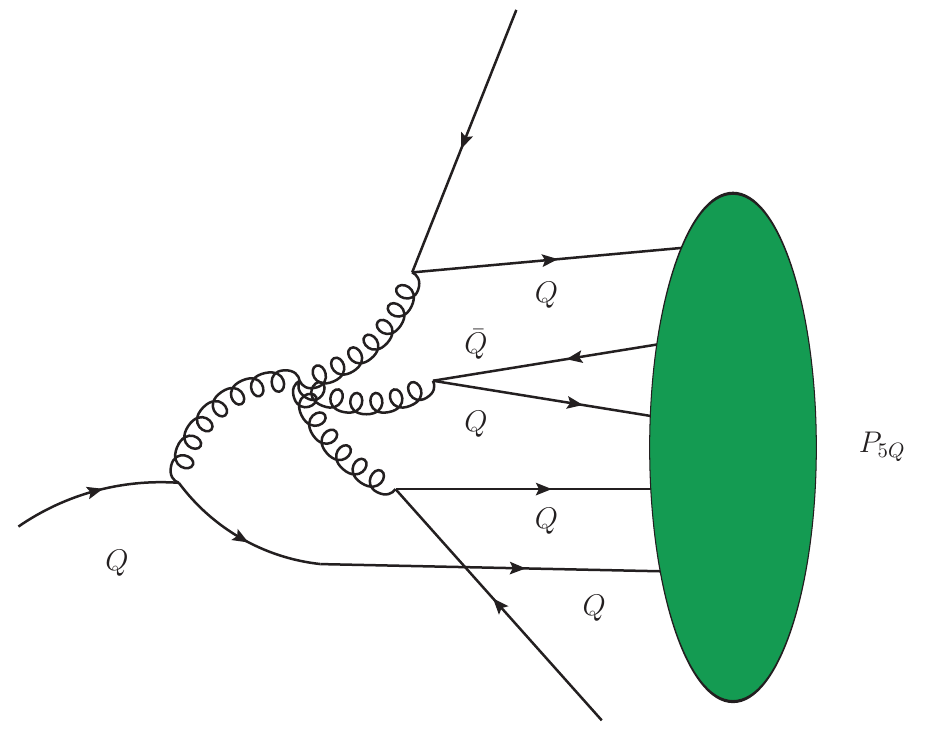}

\caption{Main leading-order diagrams for the collinear fragmentation of a constituent heavy quark into a color-singlet $S$-wave $\PQQ$ pentaquark at the initial scale, illustrated within the direct multiquark framework. The green ovals represent the nonperturbative hadronization contribution to the associated FFs. Visualizations were generated using {\tt JaxoDraw~2.0}~\cite{Binosi:2008ig}.}
\label{fig:PQQ_FF_direct}
\end{figure*}

Our construction of the initial-scale input for constituent heavy-quark fragmentation into $S$-wave color-singlet pentaquarks in the \emph{direct multiquark} scenario (Fig.~\ref{fig:PQQ_FF_direct}) builds upon the approach of Ref.~\cite{Farashaeian:2024son}, based on a spin-aware Suzuki framework~\cite{Suzuki:1977km} incorporating transverse-momentum effects.
The collinear limit is recovered by neglecting the relative motion of the bound-state constituents~\cite{Lepage:1980fj}.

This setup parallels the NRQCD factorization picture, where perturbative splittings produce the heavy constituents above threshold, followed by nonperturbative binding.
In analogy, the production amplitude is constructed from a sequence of perturbative emissions and convoluted with a bound-state wave function according to the Suzuki prescription.

The channels shown in Fig.~\ref{fig:PQQ_FF_direct} correspond to the $[Q \to (Q \bar{Q} Q Q Q) + \bar{Q} \bar{Q}]$ subprocess, namely the direct $[Q \to \PQQ]$ SDC.
Charge-conjugated configurations yield the corresponding $[\bar{Q} \to \bPQQ]$ channel.
In this work, we assume symmetry between $\PQQ$ and $\bPQQ$ production, consistently with observables inclusive over particle-antiparticle emission (see, \emph{e.g.}, Ref.~\cite{Bertone:2018ecm}).

Using {\symJethad} interfaced with {\FeynCalc}~\cite{Mertig:1990an}, we derive the explicit form of the $[Q,\bar{Q} \to \PQQ]$ initial-scale FF in the direct scenario,
\begin{equation}
\begin{split}
 \label{PQQ_FF_initial-scale_Q_direct}
 D^{\PQQ}_{Q,\,{\rm [direct]}}(z,\mu_{F,0}) \,=\,
 {\cal N}_{P,\,{\rm [direct]}}^{(Q)} \,
 (1-z)^4 z^4
 \,
 {\cal R}_{P/Q}^2
 \,
 \frac{{\cal S}_{P,\,{\rm [direct]}}^{(Q)}(z; {\cal R}_{q_T/Q}) }{{\cal T}_{P,\,{\rm [direct]}}^{(Q)}(z; {\cal R}_{q_T/Q}, {\cal R}_{P/Q})}
 \;,
\end{split}
\end{equation}
with ${\cal R}_{P/Q} = m_{\PQQ}/m_Q$ and ${\cal R}_{q_T/Q} \equiv \sqrt{\vqTTa}/m_Q$.
We set $m{\PQQ} = 5m_Q$, with $m_Q \equiv m_c = 1.5$~GeV.

The normalization factor reads as
\begin{equation}
 \label{PQQ_FF_initial-scale_Q_N_direct}
 {\cal N}_{P,\,{\rm [direct]}}^{(Q)} \, = \,
 \left\{ 320 \sqrt{5} \pi^2 \, f_{\cal B} \, C_F \big[ \alpha_s\big(\mu_{F,0}^{\rm [direct]}\big) \big]^3 \right\}^2
 \,.
\end{equation}
while the numerator and denominator are 
\begin{equation}
\label{PQQ_FF_initial-scale_Q_num_direct}
\begin{split}
 {\cal S}_{P,\,{\rm [direct]}}^{(Q)}(z; {\cal R}_{q_T/Q}) 
\,&=\,
 \sum\limits_{k=0}^9 \, z^{2k} \, \gamma_{P,\,{\rm [direct]}}^{(Q)}(z; k) 
 \left({\cal R}_{q_T/Q}\right)^{2k}
\end{split}
\end{equation}
and
\begin{equation}
\label{PQQ_FF_initial-scale_Q_den_direct}
\begin{split}
 {\cal T}_{P,\,{\rm [direct]}}^{(Q)}(z; {\cal R}_{q_T/Q}, {\cal R}_{P/Q}) 
\,&=\, 
 [{\cal R}_{q_T/Q}^2 \, z^2 + (5-3 z)^2]^7 \\[0.20cm]
\,&\times\,
 [{\cal R}_{q_T/Q}^2 \, 3z^2 - 13z^2-50 z+75]^2 \\[0.20cm]
\,&\times\, 
 [{\cal R}_{q_T/Q}^2 \, (z+2) z^2 + z^3-4 z^2-35 z+50]^2 \\[0.20cm]
\,&\times\, 
 [{\cal R}_{q_T/Q}^2 (2z-1) + (1-z)({\cal R}_{P/Q}^2-5) + 4]^2
  \;.
\end{split}
\end{equation}
respectively.
The $\gamma_{P,\,{\rm [direct]}}^{(Q)}(z; k)$ entering Eq.~\eqref{PQQ_FF_initial-scale_Q_num_direct} are detailed in Appendix~\hyperlink{app:A}{A}.

Compared to Ref.~\cite{Farashaeian:2024son}, our result differs in the explicit determination of ${\cal N}_{P,\,{\rm [direct]}}^{(Q)}$ and in a revised treatment of the $\vqTTa$ parameter.
Within the Suzuki framework, $\vqTTa$ controls the transverse-momentum structure of the bound state and can be interpreted as an effective quantity entering the F-NPWF.
Following the strategy adopted in our previous {\tt 1.0} releases~\cite{Celiberto:2025ipt}, we treat $\vqTTa$ as a phenomenological parameter governing the shape of the FF, and we will estimate the associated F-NPWF uncertainties through its variation (Sec.~\ref{ssec:FFs-PQ5Q11}).

In earlier studies on all-heavy tetraquarks~\cite{Celiberto:2024mab}, $\vqTTa$ was tuned to ensure $\langle z \rangle \gtrsim 0.4$ and a comparable normalization between quark- and gluon-initiated channels.
Adopting a similar criterion for pentacharm, we select~\cite{Celiberto:2025ipt}
\begin{equation}
 \label{eq:vqTTa_PQQ}
 \vqTTa_{\PQQ} \equiv 90 \, \text{GeV}^2 \;,
 \end{equation}
which approximately satisfies
\begin{equation}
 \label{eq:vqTTa_PQQ_TQQ}
 \sqrt{\vqTTa_{\PQQ}} \, \approx \, \frac{5}{4} \, \sqrt{\vqTTa_{\TQQ}} \;.
\end{equation}

Although this scale lies well above the typical hadronic regime, its role should be regarded as effective rather than strictly associated with a soft nonperturbative scale.
In fully heavy systems, the absence of light constituents prevents the emergence of a characteristic scale set by $\LQCD$, while the presence of multiple heavy quarks suggests a broader intrinsic momentum distribution within the bound state.
In this context, $\vqTTa$ parametrizes the intrinsic momentum spread of the bound state and should not be interpreted as a direct measure of nonperturbative dynamics at low scales.
Rather, it effectively encodes nonperturbative modulation effects associated with the formation of a fully heavy Fock state, where the relevant dynamical scale is set by the interplay between heavy-quark masses and binding interactions, rather than by $\LQCD$ alone.

This interpretation is consistent with early studies of heavy-flavor fragmentation~\cite{Suzuki:1977km,Bjorken:1977md}, where the large-$z$ behavior is driven by the heavy-quark mass.
However, for fully heavy systems such as ordinary and exotic $\TQQ$ or $\PQQ$ states, the lack of a light degree of freedom prevents a purely kinematic determination of the FF peak, thus motivating a phenomenological treatment.

\subsection{Diquark fragmentation mode}
\label{ssec:FFs-Q-diquark}

\begin{figure*}[!t]
\centering
\includegraphics[width=0.475\textwidth]{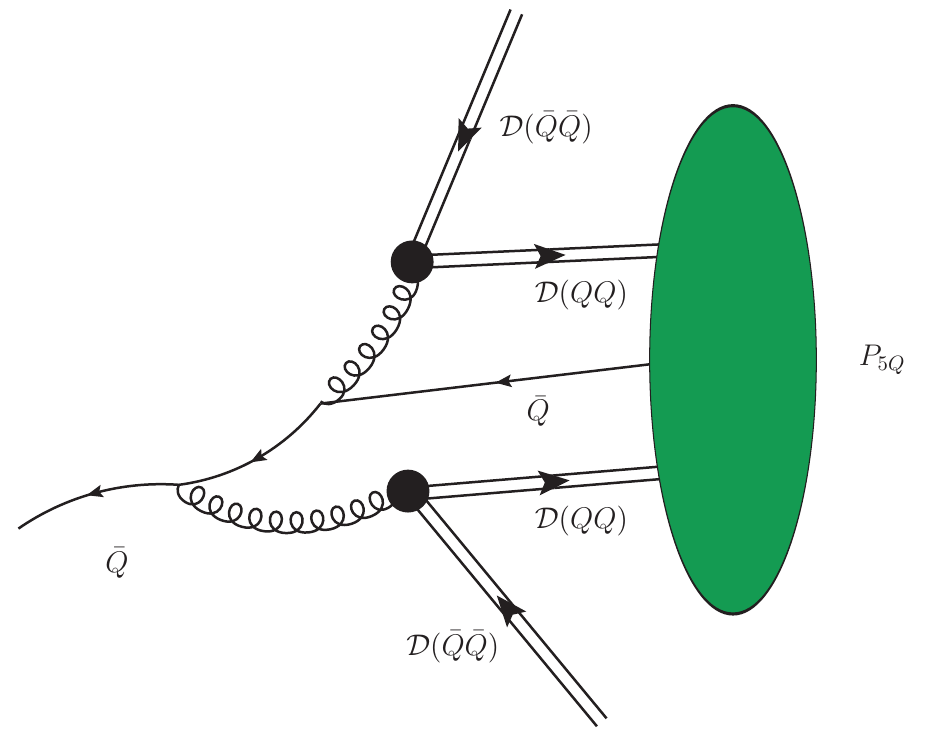}
\hspace{0.40cm}
\includegraphics[width=0.475\textwidth]{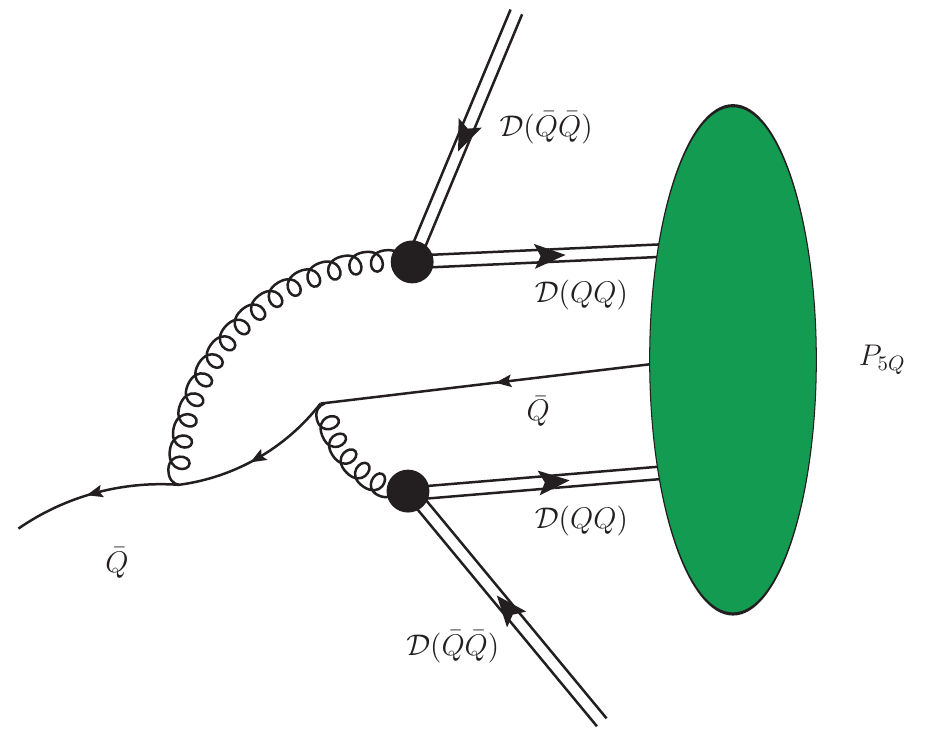}

\vspace{0.20cm}

\includegraphics[width=0.475\textwidth]{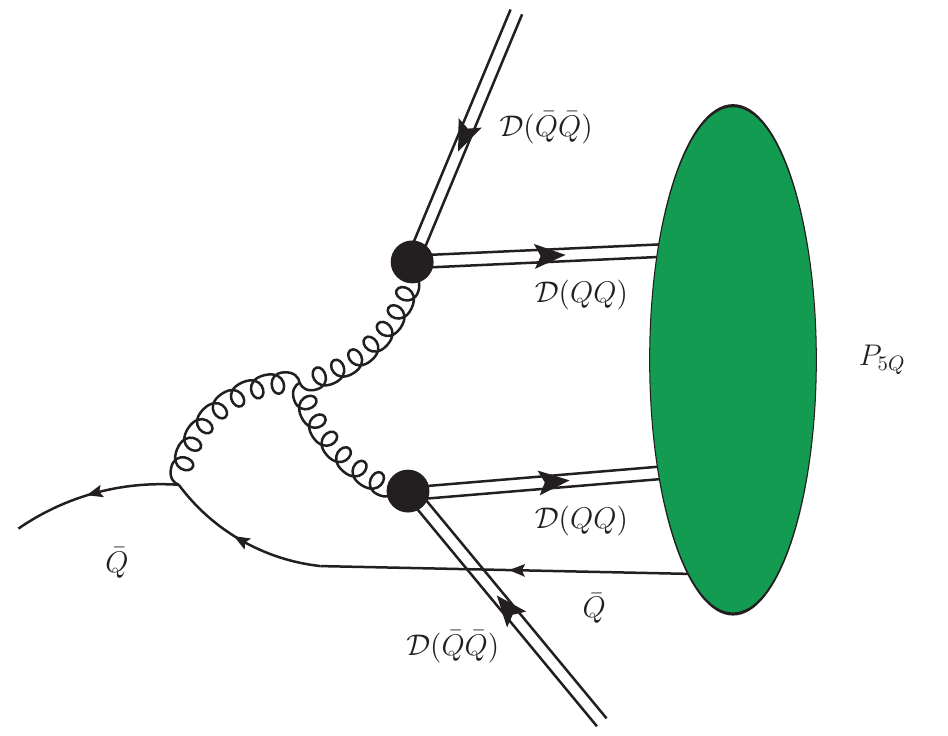}

\caption{Main leading-order diagrams for the collinear fragmentation of a constituent heavy antiquark into a color-singlet $S$-wave $\PQQ$ pentaquark at the starting scale, illustrated within the scalar-diquark framework. Double lines denote heavy-diquark states, ${\cal D}(QQ)$ or ${\cal D}(\bar{Q}\bar{Q})$, while black bullets indicate effective gluon-diquark-antidiquark vertices. The green ovals represent the nonperturbative hadronization contribution to the associated FFs. Visualizations were generated using {\tt JaxoDraw~2.0}~\cite{Binosi:2008ig}.}
\label{fig:PQQ_FF_diquark}
\end{figure*}

The quark-diquark picture assumes that two quarks can form a colored correlated subsystem, referred to as a diquark~\cite{Gell-Mann:1964ewy}.
This framework has been widely employed in hadron spectroscopy and in the description of baryon production, particularly in the heavy-flavor sector (see Refs.~\cite{Maiani:2004vq,Jaffe:2003sg,Guo:2013xga}).
Diquarks can appear in scalar (spin-$0$) or axial-vector (spin-$1$) configurations; analogous spectator systems are widely used to model spin-dependent quark~\cite{Bacchetta:2008af} and gluon densities~\cite{Bacchetta:2020vty,Bacchetta:2024fci} in the proton, with their composite nature encoded through phenomenological form factors. 
Scalar states involve a single form factor, while axial-vector ones require a more involved structure.
Early applications to baryon fragmentation are given in Refs.~\cite{Ma:2001ri,Yang:2002gh,Adamov:1997yk,MoosaviNejad:2017rvi,Delpasand:2019xpk}, while tetraquark and pentaquark systems have been explored within diquark-based approaches in Refs.~\cite{Faustov:2020qfm,Maiani:2015vwa}.

Following the strategy adopted in our previous {\tt 1.0} releases~\cite{Celiberto:2025ipt}, we model the initial-scale input for charm fragmentation into $S$-wave color-singlet pentacharms in the \emph{diquark-quark-diquark} configuration (Fig.~\ref{fig:PQQ_FF_diquark}) based on Ref.~\cite{Farashaeian:2024cpd}.
As in the direct scenario, the Suzuki framework~\cite{Suzuki:1977km} is employed, but the leading Fock state is taken as $| {\cal D}(QQ) \, \bar{Q} \, {\cal D}(QQ) \rangle$, where ${\cal D}(QQ)$ is a colored heavy diquark.
At LO, the transition $[\bar{Q} \,\to\, ({\cal D}(QQ) \, \bar{Q} \, {\cal D}(QQ)) \,+\, {\cal D}(\bar{Q}\bar{Q}) \, {\cal D}(\bar{Q}\bar{Q})]$ is described by the diagram classes shown in Fig.~\ref{fig:PQQ_FF_diquark}.

The channels in Fig.~\ref{fig:PQQ_FF_diquark} correspond to the diquark-based $[\bar{Q} \to \PQQ]$ SDC, with charge-conjugated configurations yielding the $[Q \to \bPQQ]$ channel.
As in the direct case, we assume symmetry between pentaquark and antipentaquark production, working with fully symmetric $Q$ and $\bar{Q}$ fragmentation channels.

Within this setup, the diquark-quark-diquark structure leads to more compact analytic expressions than the direct multiquark case.
Using {\symJethad} in combination with {\FeynCalc}, we reproduce the $[Q,\bar{Q} \to \PQQ]$ initial-scale {\tt PQ5Q1.1} FF in the diquark scenario (see Sec.~4 of Ref.~\cite{Farashaeian:2024cpd}),
\begin{equation}
\begin{split}
 \label{PQQ_FF_initial-scale_Q_diquark}
 D^{\PQQ}_{Q,\,{\rm [diquark]}}(z,\mu_{F,0}) \,=\,
 {\cal N}_{P,\,{\rm [diquark]}}^{(Q)} \,
 \left[\frac{z^2(1 - z)}{z + 2}\right]^2
 \,
 \frac{{\cal S}_{P,\,{\rm [diquark]}}^{(Q)}(z; {\cal R}_{q_T/Q}) }{{\cal T}_{P,\,{\rm [diquark]}}^{(Q)}(z; {\cal R}_{q_T/Q})}
 \;.
\end{split}
\end{equation}
where ${\cal R}_{q_T/Q} = \sqrt{\vqTTa}/m_Q$\,.
\begin{equation}
 \label{PQQ_FF_initial-scale_Q_N_diquark}
 {\cal N}_{P,\,{\rm [diquark]}}^{(Q)} \, = \,
 \left[\frac{{\cal V}_{P,\,\rm [diquark]}^{(g{\cal D}\bar{{\cal D}})}}{m_Q}\right]^4
 \!
 \left\{ \frac{625 \pi^2}{6 \sqrt{2}} \, f_{\cal B} \, C_F \big[ \alpha_s\big(\mu_{F,0}^{\rm [diquark]}\big) \big]^2 \right\}^2
 \,,
\end{equation}
while the numerator and denominator explicitly read as
\begin{equation}
\label{PQQ_FF_initial-scale_Q_num_diquark}
\begin{split}
 {\cal S}_{P,\,{\rm [diquark]}}^{(Q)}(z; {\cal R}_{q_T/Q}) 
\,&=\,
 \left\{ 64z^4 + 356z^3 - 99z^2 - 2128z + 1540 + \frac{672}{z} \right. \\[0.20cm]
\,&+\, 
 225 \left[ \frac{128z(1 - z)^6}{[{\cal R}_{q_T/Q}^2 \, z^2 + (5 - z)^2]^2} - \frac{45(z + 2)^2(8z^2 + 60z - 33 - 60/z)}{[{\cal R}_{q_T/Q}^2 \, z^2 + z^2 - 10z + 100]^2} \right] \\[0.20cm]
\,&+\,
 30 \left[ \frac{16(1 - z)^4 (z^2 - 20z - 35)}{{\cal R}_{q_T/Q}^2 \, z^2 + (5 - z)^2} \right] \\[0.20cm]
\,&+\,
 \left. \frac{(z + 2)(100z^4 + 256z^3 - 588z^2 + 1859z - 922 - 1380/z)}{{\cal R}_{q_T/Q}^2 \, z^2 + z^2 - 10z + 100} \right\}
\end{split}
\end{equation}
and
\begin{equation}
\label{PQQ_FF_initial-scale_Q_den_diquark}
\begin{split}
 {\cal T}_{P,\,{\rm [diquark]}}^{(Q)}(z; {\cal R}_{q_T/Q}) 
  \,&=\, 
 [{\cal R}_{q_T/Q}^2 \, z^2 + (5-z)^2]^2
  \;.
\end{split}
\end{equation}
The quantity ${\cal V}_{P,\,\rm [diquark]}^{(g{\cal D}\bar{{\cal D}})}$ in Eq.~\eqref{PQQ_FF_initial-scale_Q_N_diquark} is the form factor part of the gluon-diquark-antidiquark effective vertex (the black bullet in Fig.~\ref{fig:PQQ_FF_diquark}). 

In our implementation, the calculation of Ref.~\cite{Farashaeian:2024cpd} is used as a proxy for the diquark channel, retaining scalar diquarks only and neglecting pseudovector contributions.
We improve upon that setup in three aspects.
First, the normalization factor in Eq.~\eqref{PQQ_FF_initial-scale_Q_diquark} is computed explicitly at the initial scale $\mu_{F,0}^{\rm [diquark]} = 9 m_Q$, rather than fixed by normalization conditions.
Second, the $\vqTTa$ parameter is determined consistently with the direct scenario (see Eq.\eqref{eq:vqTTa_PQQ}).
Third, to better isolate the impact of diquark dynamics, we set the $(g{\cal D}\bar{{\cal D}})$ form factor to ${\cal V}_{P,\,\rm [diquark]}^{(g{\cal D}\bar{{\cal D}})} = 1~\text{GeV}$, instead of $5~\text{GeV}$ as in Ref.~\cite{Farashaeian:2024cpd}.

\subsection{The {\tt PQ5Q1.1} multimodal functions}
\label{ssec:FFs-PQ5Q11}

The construction of the {\tt PQ5Q1.1} fragmentation functions for $\PQc$ states is completed by consistently evolving the initial-scale inputs through DGLAP dynamics.  
This step connects the hadron-structure-driven modeling introduced in the previous sections to the perturbative description of fragmentation at higher energy scales.

From kinematic considerations, the minimal invariant mass for the $[c \to (c \bar{c} c c c) + \bar{c} \bar{c}]$ splitting in the direct multicharm scenario is $\mu_{F,0}^{\rm [direct]} = 7 m_c$ (see Fig.~\ref{fig:PQQ_FF_direct}).
Conversely, the minimal invariant mass for the $[\bar{c} \to (c \bar{c} c c c) + \bar{c} \bar{c} \bar{c} \bar{c}]$ splitting in the scalar-dicharm mode is $\mu_{F,0}^{\rm [diquark]} = 9 m_c$ (see Fig.~\ref{fig:PQQ_FF_diquark}).
These scales define the natural starting points for the fragmentation evolution in the two configurations.

A consistent treatment of heavy-hadron fragmentation requires matching hadronic-structure inputs with collinear factorization, ensuring a proper implementation of evolution thresholds across all parton channels.  
To achieve this, the \emph{heavy-flavor nonrelativistic evolution} ({\HFNRevo}) framework was introduced~\cite{Celiberto:2025euy,Celiberto:2024mex,Celiberto:2024bxu,Celiberto:2024rxa,Celiberto:2025xvy,Celiberto:2026rzi,Celiberto:2026zss}.  
Originally designed to describe the DGLAP evolution of fragmentation functions derived from nonrelativistic inputs, and later extended to more general initial-scale models, {\HFNRevo} is structured around three key elements: \emph{interpretation}, \emph{evolution}, and \emph{uncertainties}.

Concerning the evolution stage, {\HFNRevo} can be viewed as a two-step procedure.  
An initial semianalytic \emph{expanded} and \emph{decoupled} evolution ({\tt EDevo}) ensures a consistent treatment of threshold effects across parton channels.  
This is followed by a fully numerical \emph{all-order} evolution step ({\tt AOevo}), which implements the complete DGLAP dynamics.

In the context of exotic hadrons, {\HFNRevo} has been successfully applied to the description of fully heavy tetraquark fragmentation, combining gluon and constituent heavy-quark initial-scale inputs (see, e.g., the {\tt TQ4Q1.0} sets~\cite{Celiberto:2024mab} and their {\tt 1.1}~\cite{Celiberto:2024beg,Celiberto:2025dfe,Celiberto:2025ziy} and {\tt 2.0}~\cite{Celiberto:2026kks} extensions, as well as the $\Omega_{3c}$ {\tt OMG3Q1.0} functions~\cite{Celiberto:2025ogy}).

When only a single parton channel is available at the initial scale, a simplified realization of {\HFNRevo} is sufficient.  
This occurs, for instance, in the {\tt TQHL1.0} framework~\cite{Celiberto:2023rzw} (and its {\tt 1.1} update~\cite{Celiberto:2024beg}), where only the $[Q,\bar{Q} \to \XQq]$ channel is modeled.  
In such cases, the {\tt EDevo} step is bypassed, and the evolution proceeds directly through the numerical {\tt AOevo} stage.

A similar situation applies to $\PQc$ fragmentation.  
At present, the only available initial-scale input corresponds to the constituent heavy-quark channel, namely the charm.  
Therefore, in constructing the {\tt PQ5Q1.1} set, we directly evolve the $[c,\bar{c} \to \PQc]$ inputs through the {\tt AOevo} procedure, while all other partonic contributions are dynamically generated by the evolution.

Within this setup, the charm-quark mass $m_c$ enters explicitly in the initial condition for the $[c,\bar{c} \to \PQc]$ channel, through the multimodal, mass-dependent construction discussed in the previous sections.  
This reflects the fact that the formation of a $\PQc$ state occurs close to the heavy-quark threshold, where finite-mass effects are relevant.  
As a consequence, the subsequent evolution is performed in a fixed-flavor scheme with $n_f = 5$, without introducing additional threshold crossings.

The DGLAP evolution is implemented within the ZM-VFNS, where all partons, including charm and bottom, are treated as massless degrees of freedom.  
This choice is standard in heavy-hadron fragmentation studies (see, \emph{e.g.}, Ref.~\cite{Cacciari:2024kaa}), and is justified by the fact that mass effects become negligible at the scales relevant for the evolution.  
In this context, the ZM-VFNS label refers exclusively to the massless treatment in the evolution stage, while mass effects are retained in the initial-scale modeling.

One limitation of the present approach is that initial-scale inputs for light partons and the bottom quark are not explicitly included, and only emerge through evolution at higher scales.  
However, existing studies indicate that such contributions are negligible at the initial scale for analogous systems, such as doubly heavy tetraquarks~\cite{Nejad:2021mmp} and vector quarkonia~\cite{Celiberto:2022dyf}.  
Extending the present framework to include additional initial channels will be the subject of future investigations.

A key feature of the {\tt PQ5Q1.1} framework is the explicit construction of a replica ensemble designed to encode the dominant sources of theoretical uncertainty in a coherent and systematic way.  
Rather than treating perturbative and nonperturbative effects separately at the level of observables, both are incorporated directly at the level of the fragmentation functions, allowing for a unified propagation of uncertainties throughout the entire phenomenological pipeline.

On the perturbative side, uncertainties are estimated through F-MHOUs.  
These are implemented by varying the scale that sets the starting point of the numerical {\tt AOevo} evolution, namely $\mu_{F,0}^{\rm [direct]} = 7 m_c$ in the direct mode and $\mu_{F,0}^{\rm [diquark]} = 9 m_c$ in the diquark mode. 

This \emph{evolution-ready} scale is varied around its natural value by a multiplicative factor $K_{\mu}$, ranging from $1/2$ to $2$, thus probing the sensitivity to subleading perturbative contributions beyond the nominal accuracy.   
Our procedure is fully consistent with the {\HFNRevo} strategy, where the initial-scale modeling and the subsequent DGLAP evolution are treated as a single, continuous framework.

In addition, nonperturbative uncertainties are accounted for through variations of the transverse-momentum parameter $\vqTTa$, which controls the shape of the nonperturbative wave function in the Suzuki-inspired construction.  
Although $\vqTTa$ does not correspond to a soft hadronic scale in fully heavy systems, it effectively parametrizes the internal momentum distribution of the multiquark bound state, thus providing a controlled handle on the modeling of bound-state formation.  
The resulting F-NPWF variations probe the sensitivity of the FFs to the underlying hadron-structure assumptions.

These perturbative (F-MHOU) and nonperturbative (F-NPWF) effects are combined within a replicalike structure, where each member of the ensemble corresponds to a specific choice of energy-scale configuration and transverse-momentum parameter.  
In practice, this leads to a discrete set of FF replicas that encode the combined uncertainty envelope, which can be directly propagated to collider observables.  

We stress that these replicas do not constitute a Monte Carlo statistical ensemble, as in modern PDF determinations, but rather a structured scan over physically motivated parameter variations.  
Their purpose is not to reconstruct a probability distribution, but to provide a controlled and transparent sampling of the dominant theoretical uncertainties affecting the fragmentation process. 

This approach is particularly advantageous in the present context, where uncertainties originate from modeling assumptions (NPWF) and missing higher-order effects (F-MHOUs), for which no statistically rigorous prior is available.  
By organizing these variations into a replicalike format, one retains the practical benefits of replica-based frameworks---such as straightforward uncertainty propagation, envelope construction, and compatibility with standard analysis tools---while maintaining a clear interpretation in terms of theory-driven variations.  

The explicit mapping between replica indices and parameter configurations is provided in Appendix~\hyperlink{app:B}{B}, where the full structure of the ensemble is detailed.

\begin{figure*}[!t]
\centering

   \hspace{-0.00cm}
   \includegraphics[scale=0.410,clip]{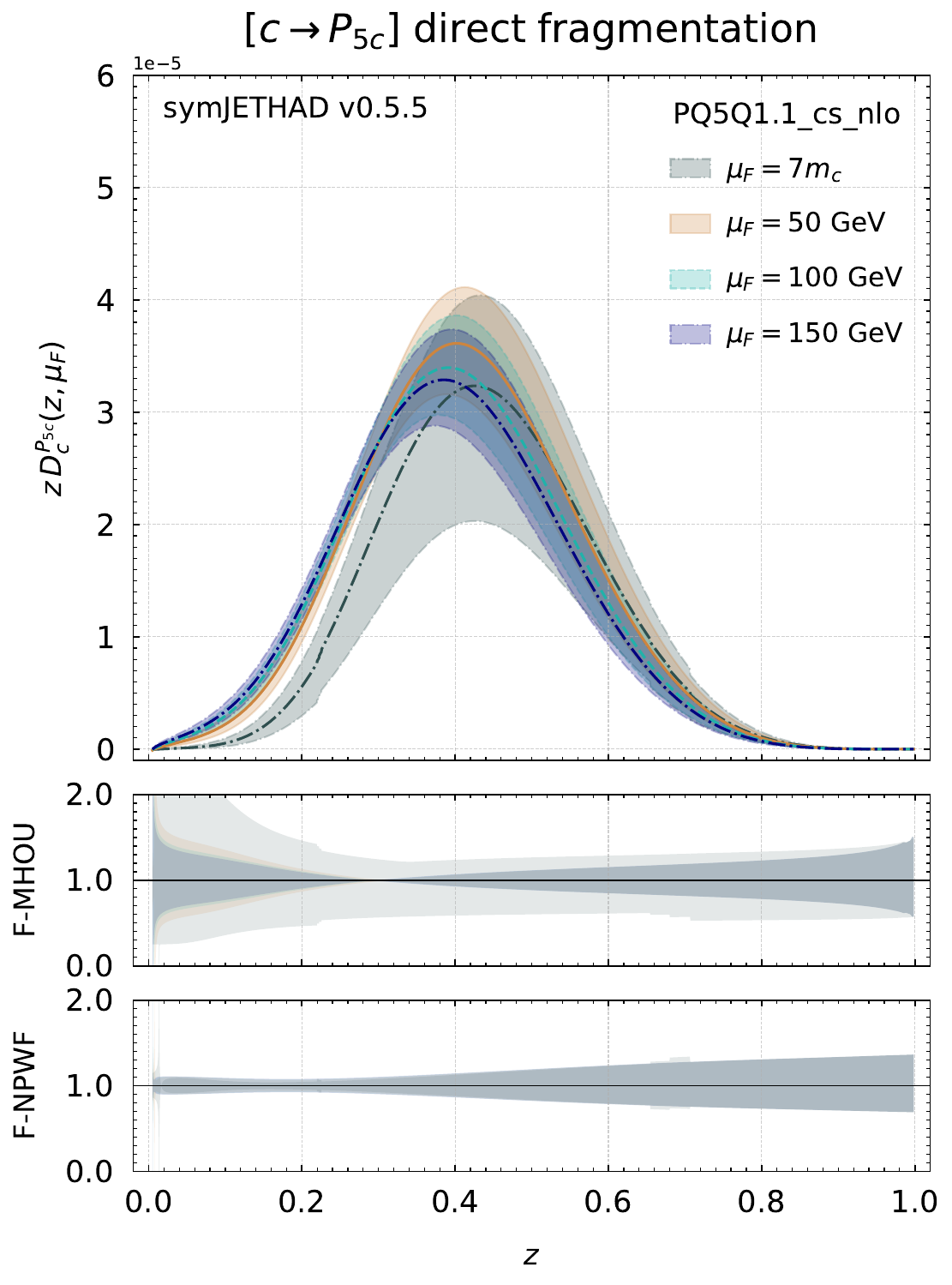}
   \hspace{0.90cm}
   \includegraphics[scale=0.410,clip]{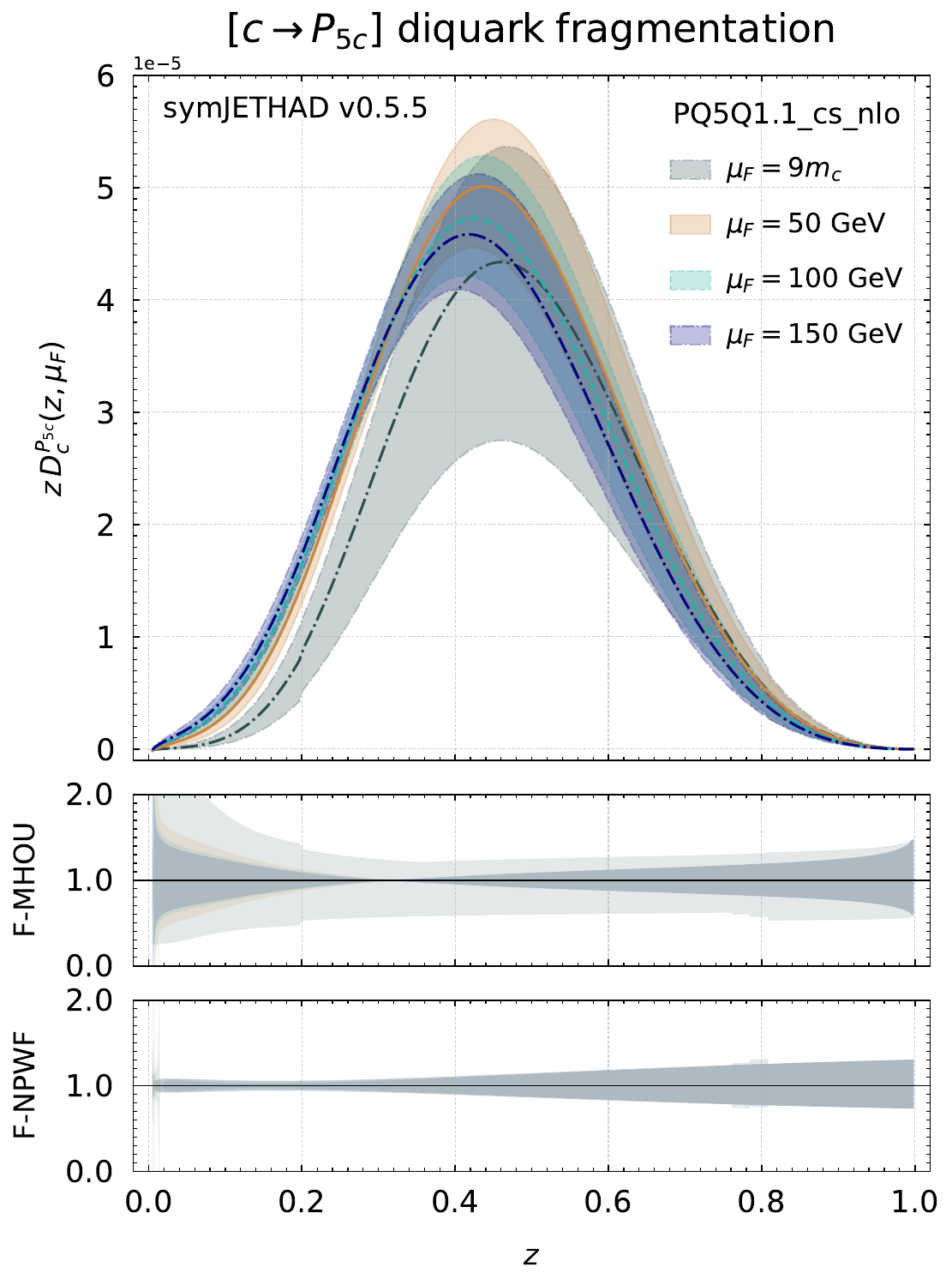}

\caption{
Momentum-fraction dependence of {\tt PQ5Q1.1} functions describing the collinear fragmentation of (anti)charm quarks into $\PQc$ pentaquarks, for direct (left) and scalar-diquark (right) initial-scale inputs. 
Shaded bands in the main panels denote the total uncertainty, obtained by combining F-MHOU and F-NPWF variations. 
The first lower panel highlights the effect of F-MHOUs, while the second isolates F-NPWF uncertainties. 
For details on the replica indexing and parameter mapping, see Table~\ref{tab:PQ5Q11_replicas} in Appendix~\protect\hyperlink{app:B}{B}.}
\label{fig:FFs-z_PQc_charm}
\end{figure*}

Figure~\ref{fig:FFs-z_PQc_charm} shows the $z$ dependence of the {\tt PQ5Q1.1} FFs for all-charm pentaquarks in the direct (left) and scalar-diquark (right) configurations. 
In the main panels, the central prediction is defined as the average over the full replica ensemble, while the shaded bands represent the total uncertainty obtained by combining perturbative F-MHOUs and nonperturbative F-NPWF effects. 
The perturbative component is estimated through correlated variations of the renormalization scale entering the initial-scale inputs and of the scale defining the starting point of the {\tt AOevo} evolution, both varied by a common factor $K_{\mu}$ between $1/2$ and $2$. 
Nonperturbative effects are instead probed by varying the transverse-momentum parameter $\vqTTa$, which controls the wave-function profile, in the range $\vqTTa = 80,\,90,\,100~\text{GeV}^2$. 
Since both sources of uncertainty are encoded within the replica ensemble, the total band directly reflects their combined impact on the FFs. 
The two lower panels disentangle these contributions, showing separately the F-MHOU and F-NPWF effects as ratios to the central curve.

To assess the role of perturbative QCD evolution, the FFs are displayed at four representative factorization scales: the evolution-ready scale, namely $\mu_{F,0}^{\rm [direct]} = 7 m_c$ in the direct mode and $\mu_{F,0}^{\rm [diquark]} = 9 m_c$ in the diquark one, together with $\mu_F = 50$, $100$, and $150$~GeV. 
In both configurations, the FFs exhibit a smooth and stable dependence on $\mu_F$, characterized by a gradual redistribution of strength toward lower $z$ values and a moderate suppression of the peak height as the scale increases. 
This behavior signals a well-controlled DGLAP evolution and confirms the robustness of the initial-scale construction in the all-charm sector.

The overall shape of the distributions is bell-like, with support concentrated in the inter\-mediate-$z$ region and a maximum located around $z \sim 0.35$--$0.45$, consistently with the expected average momentum fraction probed in semihard processes. 
A clear difference emerges between the two production modes. 
The scalar-diquark configuration displays a higher and broader peak, slightly shifted toward larger $z$, indicating a more efficient transfer of momentum to the produced pentaquark. 
Conversely, the direct configuration leads to a somewhat lower and narrower distribution, reflecting a softer fragmentation pattern. 
This qualitative hierarchy is physically consistent with the different underlying mechanisms encoded in the multimodal initial conditions.

A useful comparison can be made with the corresponding all-charm scalar tetraquark {\tt TQ4Q2.0} FFs, recently released and discussed in Ref.~\cite{Celiberto:2026kks}. 
In that case, the distributions exhibit a pronounced enhancement toward small $z$, accompanied by a mild structure at large $z$. 
This behavior originates from the NRQCD-based initial-scale inputs, where the SDCs induce characteristic endpoint enhancements. 
By contrast, the {\tt PQ5Q1.1} functions display a smooth, bell-shaped profile, with support concentrated in the intermediate-$z$ region. 
This difference reflects the distinct nature of the underlying modeling: while tetraquark FFs retain a direct imprint of perturbative splitting dynamics, the present pentaquark construction is driven by a Suzuki-type approach and by the intrinsic momentum distribution encoded in the nonperturbative wave function. 
From a physical viewpoint, this suggests that pentaquark fragmentation is more strongly constrained by bound-state formation dynamics, leading to a more localized momentum-fraction distribution.

The uncertainty bands exhibit a characteristic behavior across the $z$ range. 
They are relatively narrow around the peak region, where the FFs are largest and more stable, and widen toward the endpoints, where both perturbative and nonperturbative effects become more pronounced. 
The separation of contributions in the ancillary panels provides further insight. 
F-MHOU effects remain moderate in the bulk region and grow toward small and large $z$, as expected from the increased sensitivity on scale variations in these kinematic domains. 
On the other hand, F-NPWF uncertainties induce smoother and more systematic deformations of the distributions, with a mild impact near the peak and a progressively larger effect at higher $z$.

Overall, these results indicate that, in the all-charm sector, the {\tt PQ5Q1.1} framework delivers a stable and physically interpretable description of pentaquark fragmentation. 
Both production modes exhibit regular shapes, controlled scale evolution, and a well-defined uncertainty pattern, supporting the use of these functions for phenomenological applications.

\begin{figure*}[!t]
\centering

   \hspace{-0.00cm}
   \includegraphics[scale=0.410,clip]{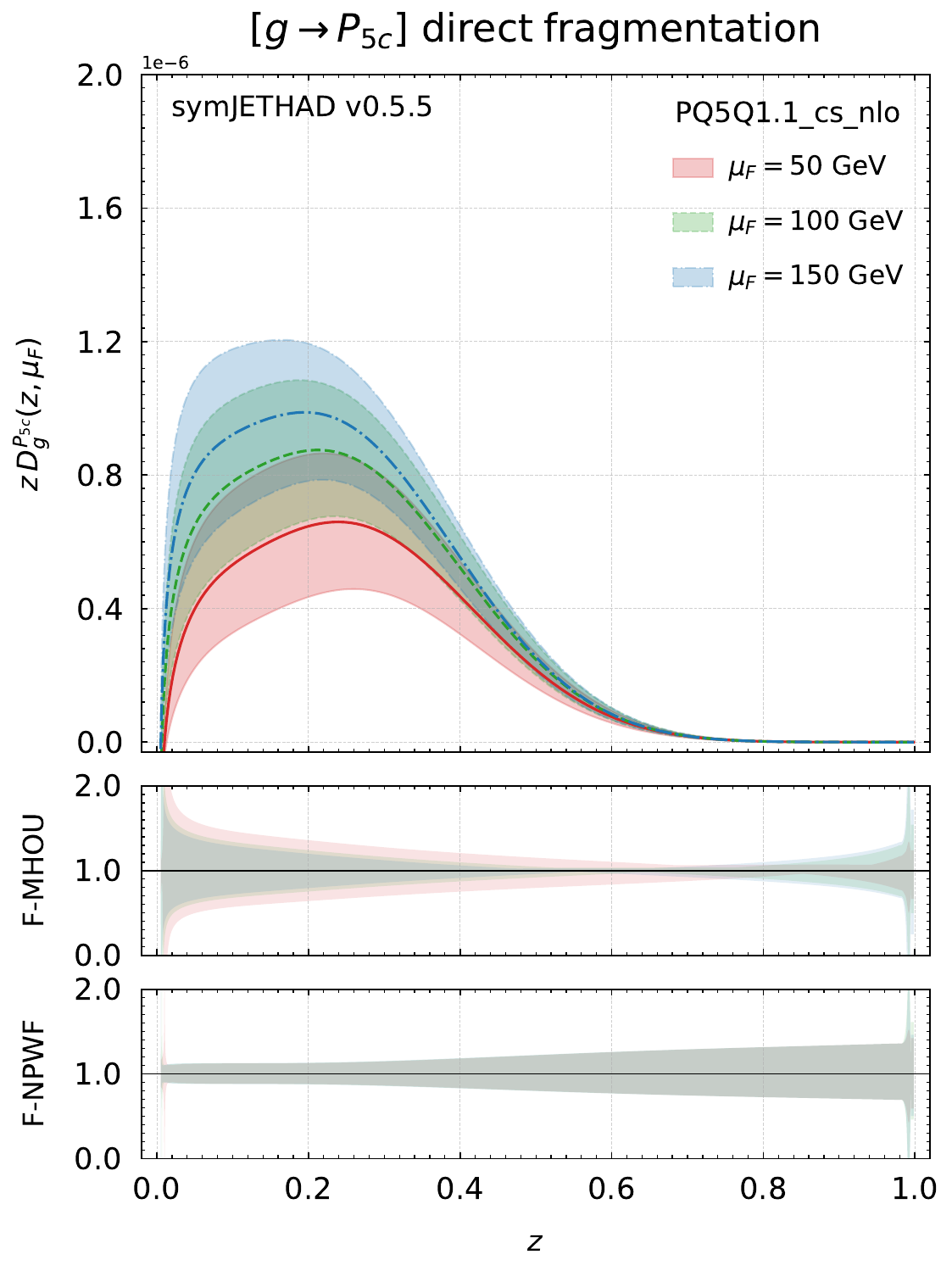}
   \hspace{0.90cm}
   \includegraphics[scale=0.410,clip]{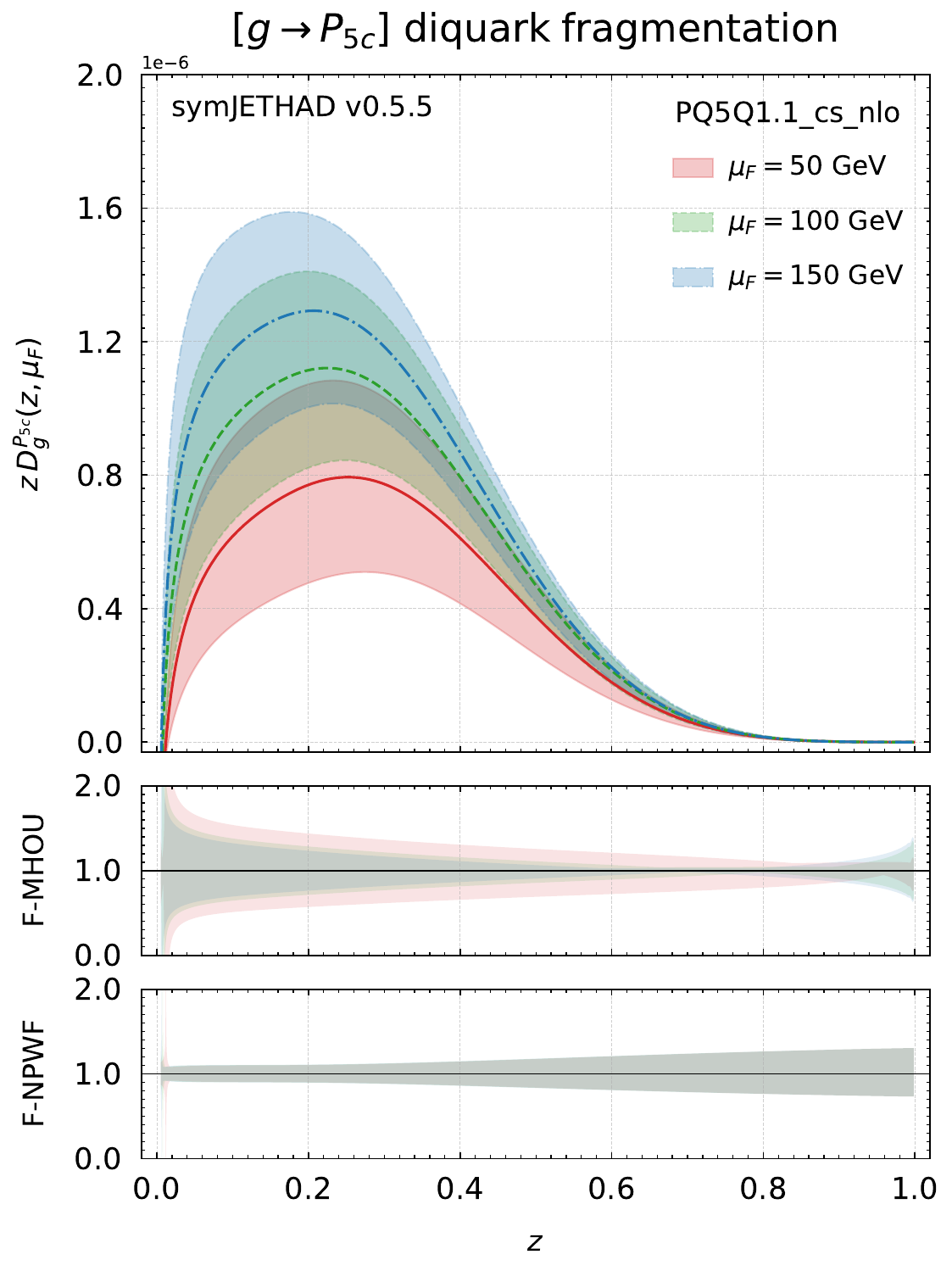}

\caption{
Momentum-fraction dependence of {\tt PQ5Q1.1} functions describing the collinear fragmentation of gluons into $\PQc$ pentaquarks, for direct (left) and scalar-diquark (right) initial-scale inputs. 
Shaded bands in the main panels denote the total uncertainty, obtained by combining F-MHOU and F-NPWF variations. 
The first lower panel highlights the effect of F-MHOUs, while the second isolates F-NPWF uncertainties. 
For details on the replica indexing and parameter mapping, see Table~\ref{tab:PQ5Q11_replicas} in Appendix~\protect\hyperlink{app:B}{B}.}
\label{fig:FFs-z_PQc_gluon}
\end{figure*}

Figure~\ref{fig:FFs-z_PQc_gluon} shows the $z$ dependence of the {\tt PQ5Q1.1} gluon FFs for all-charm pentaquarks in the direct (left) and scalar-diquark (right) configurations. 
The central prediction and the associated uncertainty bands are constructed as described in Fig.~\ref{fig:FFs-z_PQc_charm}, combining perturbative (F-MHOU) and nonperturbative (F-NPWF) effects within the replica framework. 
In contrast to the heavy-quark case, we do not display the FFs at the evolution-ready scale, since the gluon channel is not modeled at the initial scale and is therefore identically zero at $\mu_F = \mu_{F,0}^{\rm [model]}$, being dynamically generated only through DGLAP evolution.

The resulting distributions exhibit a smooth and regular behavior, with support concentrated at small-to-intermediate values of $z$ and a broad maximum around $z \sim 0.15$--$0.3$. 
The scalar-diquark mode leads to a systematically larger normalization and a slightly wider distribution with respect to the direct configuration, while preserving a similar overall shape. 
The dependence on $\mu_F$ is stable and monotonic, with increasing scales enhancing the low- and intermediate-$z$ region, consistently with the evolution-driven origin of the gluon channel.

A comparison with the corresponding gluon FFs for fully heavy tetraquarks provides additional insight. 
In the scalar $T_{4c}(0^{++})$ case (Fig.~4 of Ref.~\cite{Celiberto:2026kks}), the distributions are strongly enhanced toward small $z$, reflecting the presence of a nonvanishing gluon contribution already at leading order in the NRQCD initial conditions, which introduces endpoint-enhanced SDCs. 
By contrast, the present pentaquark FFs display a bell-shaped profile, since the gluon channel is entirely generated by evolution and does not inherit such perturbative enhancements. 
This highlights a key difference between the two systems: while tetraquark production can be directly driven by perturbative gluon fragmentation, pentaquark fragmentation in the present framework is dominated by heavy-quark–initiated dynamics, with the gluon channel emerging as a secondary, evolution-induced component.

Interestingly, a closer qualitative similarity is observed with the axial-vector $T_{4c}(1^{+-})$ channel, where the gluon contribution at leading order is suppressed by selection rules, leading to a smoother $z$ dependence (Fig.~5 of Ref.~\cite{Celiberto:2026kks}). 
In the pentaquark case, however, the shape is further modulated by the underlying charm-initiated input, which drives the evolution and imprints a characteristic intermediate-$z$ structure associated with bound-state formation dynamics.

The uncertainty pattern reflects the dynamical origin of the gluon channel. 
F-MHOU effects are more pronounced at small $z$, where the evolution is most sensitive to scale variations, while F-NPWF contributions induce smoother distortions that become increasingly relevant at larger $z$. 
Overall, the gluon FFs remain well behaved across the full $z$ range, confirming the consistency of their evolution-driven construction within the {\tt PQ5Q1.1} framework.

\begin{figure*}[!t]
\centering

   \hspace{-0.00cm}
   \includegraphics[scale=0.400,clip]{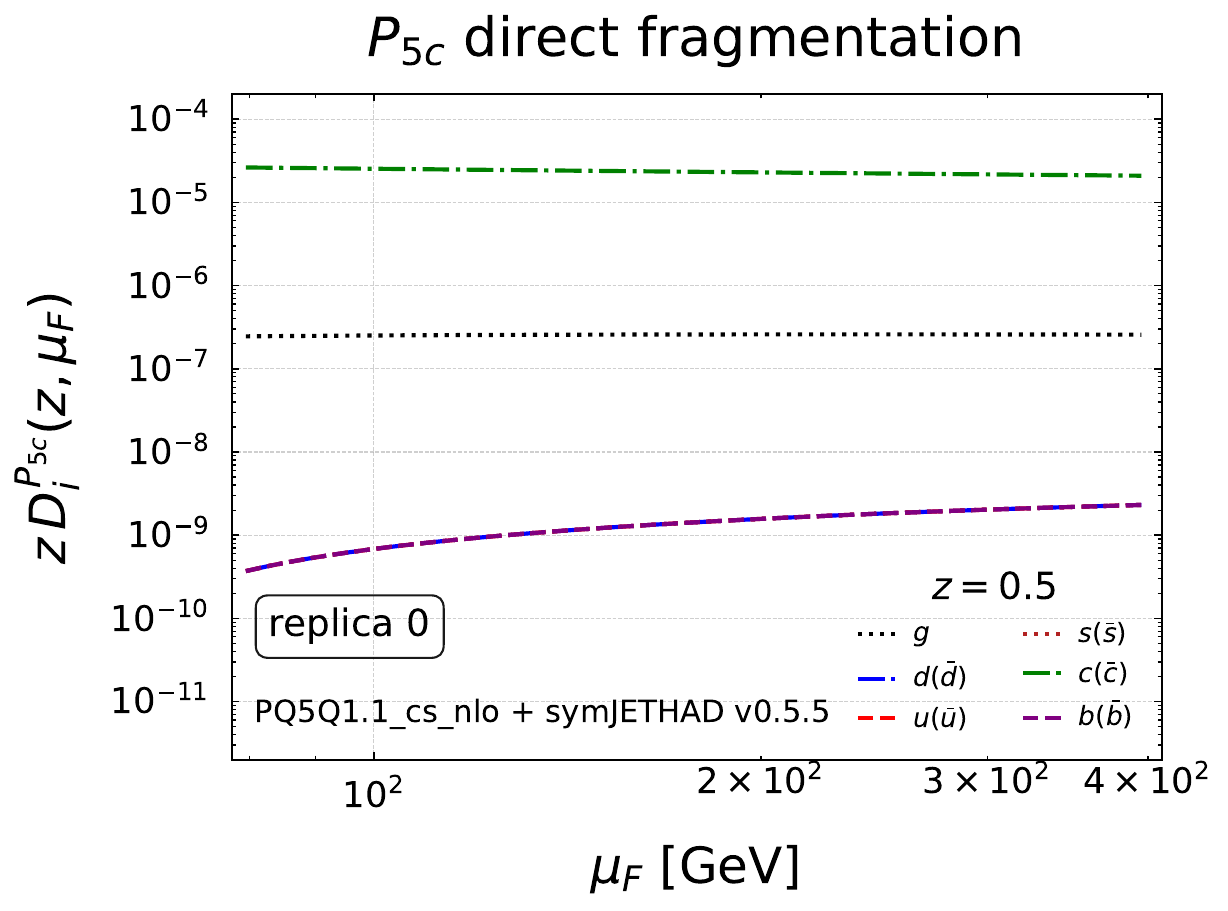}
   \hspace{-0.15cm}
   \includegraphics[scale=0.400,clip]{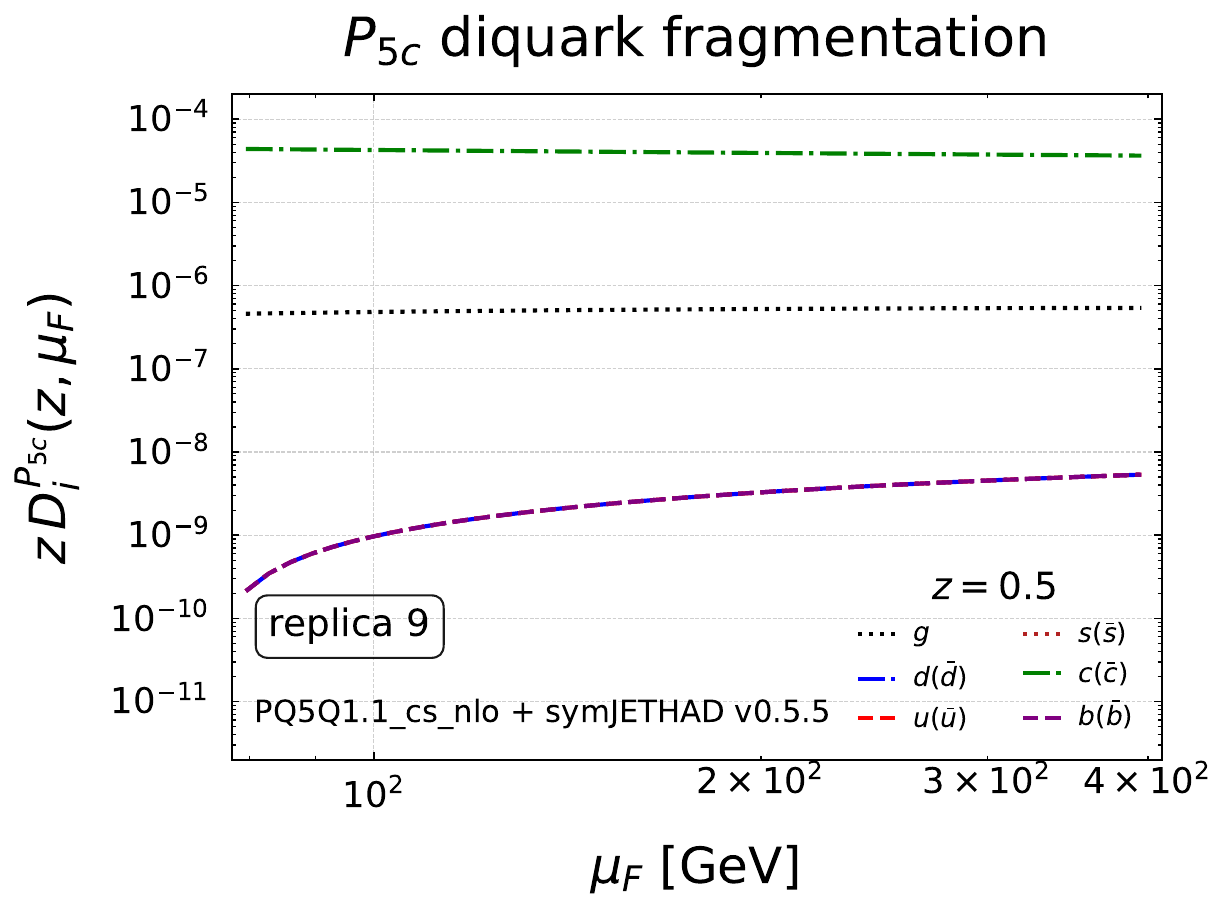}

\caption{Energy dependence of {\tt PQ5Q1.1} functions describing the collinear fragmentation of all parton species to $P_{5c}$ pentaquarks within direct (left) or scalar-diquark (right) initial-scale inputs, at $z = 0.5 \simeq \langle z \rangle$.
Central-value replicas are shown, corresponding to the default configurations for each mode (replica $0$ for the direct case and replica $9$ for the diquark one), without including F-MHOU or F-NPWF variations. 
For details on the replica indexing and parameter mapping, see Table~\ref{tab:PQ5Q11_replicas} in Appendix~\protect\hyperlink{app:B}{B}.}
\label{fig:FFs-muF_PQc}
\end{figure*}

\begin{figure*}[!t]
\centering

   \hspace{-0.00cm}
   \includegraphics[scale=0.400,clip]{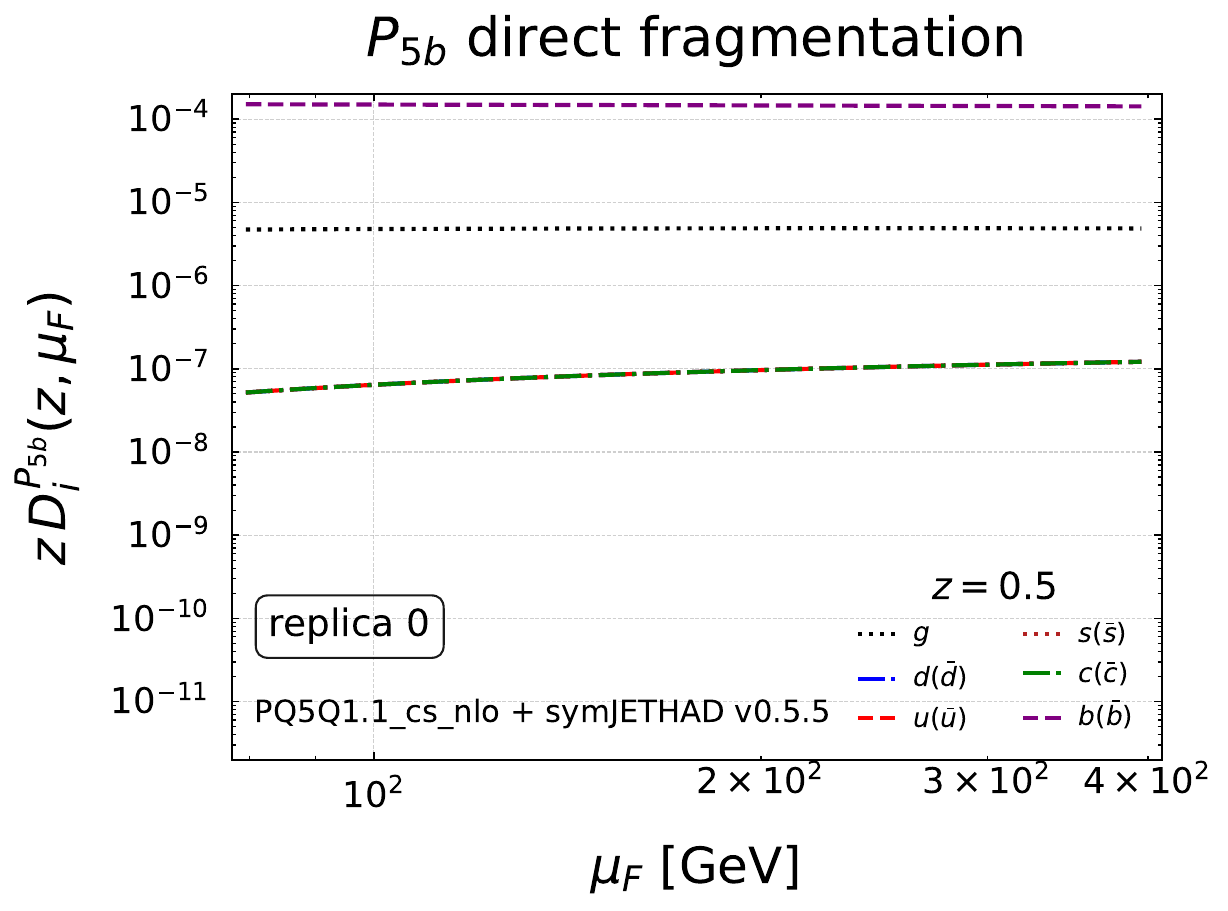}
   \hspace{-0.15cm}
   \includegraphics[scale=0.400,clip]{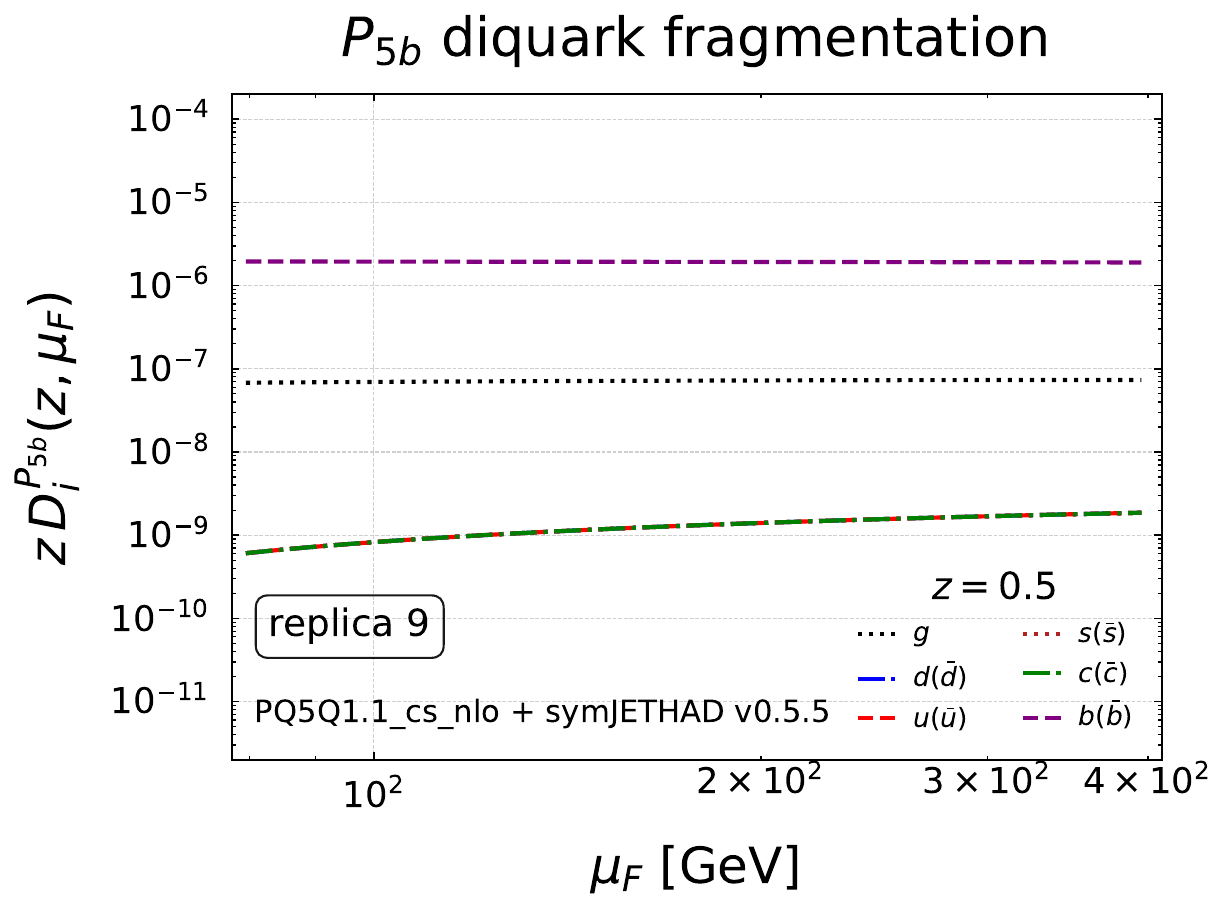}

\caption{Energy dependence of {\tt PQ5Q1.1} functions describing the collinear fragmentation of all parton species to $P_{5b}$ pentaquarks within direct (left) or scalar-diquark (right) initial-scale inputs, at $z = 0.5 \simeq \langle z \rangle$.
Central-value replicas are shown, corresponding to the default configurations for each mode (replica $0$ for the direct case and replica $9$ for the diquark one), without including F-MHOU or F-NPWF variations. 
For details on the replica indexing and parameter mapping, see Table~\ref{tab:PQ5Q11_replicas} in Appendix~\protect\hyperlink{app:B}{B}.}
\label{fig:FFs-muF_PQb}
\end{figure*}

For completeness, Fig.~\ref{fig:FFs-muF_PQc} displays the $\mu_F$ dependence of the {\tt PQ5Q1.1} FFs for all-charm pentaquarks, in both the direct (left) and scalar-diquark (right) configurations.  
For clarity, we fix the momentum fraction to $z = 0.5 \simeq \langle z \rangle$, representative of the typical kinematic region probed in semihard processes (see, \emph{e.g.}, Refs.~\cite{Celiberto:2016hae,Celiberto:2017ptm,Celiberto:2020wpk,Celiberto:2021dzy,Celiberto:2021fdp,Celiberto:2022dyf,Celiberto:2022keu,Celiberto:2022kxx,Celiberto:2024omj}). 

Central-value replicas are shown, corresponding to the default configurations for each mode (replica $0$ for the direct case and replica $9$ for the diquark one), and without including F-MHOU or F-NPWF variations. 
At variance with the momentum-fraction analysis in Figs.~\ref{fig:FFs-z_PQc_charm} and~\ref{fig:FFs-z_PQc_gluon}, where the central prediction is defined as the average over the full replica ensemble, here a single representative replica is selected for each mode to ensure a clear and reproducible comparison of the scale dependence. 
This choice does not affect the qualitative behavior of the results, as the different replicas exhibit a consistent and stable $\mu_F$ evolution pattern.

A clear hierarchy emerges, with the charm channel dominating over both light-parton and gluon contributions.  
In particular, the $c \to \PQc$ FF exceeds the gluon channel by roughly 1 to 2 orders of magnitude across the entire $\mu_F$ range explored.

Despite their smaller normalization, gluon FFs exhibit a remarkably smooth and stable dependence on the factorization scale.  
As discussed in the previous analysis, their evolution-driven origin leads to broad and regular distributions in $z$, without the endpoint-enhanced structures typical of perturbatively dominated channels.  
This behavior is a key ingredient in the so-called \emph{natural stability} of high-energy observables, whereby the gluon contribution acts as a stabilizing component against large perturbative corrections. 
Such an effect has been observed in semi-inclusive heavy-flavor production, both for singly~\cite{Celiberto:2021dzy,Celiberto:2021fdp} and multiply~\cite{Celiberto:2022dyf,Celiberto:2022keu,Celiberto:2023rzw} heavy systems.

In the present case, the interplay between the dominant charm channel and the smoother gluon component provides a balanced contribution to the cross section, improving the overall stability of resummed predictions.  
This mechanism will play a central role in the phenomenological analysis presented in Sec.~\ref{sec:phenomenology} (see also Sec.~\ref{ssec:resummation}).

Although the present analysis is primarily focused on all-charm pentaquarks, we report in Fig.~\ref{fig:FFs-muF_PQb} a representative subset of our findings for the all-bottom case, showing the $\mu_F$ dependence of the {\tt PQ5Q1.1} FFs in direct (left) and scalar-diquark (right) configurations at fixed $z = 0.5 \simeq \langle z \rangle$. 
The results are displayed for central-value replicas, without including F-MHOU or F-NPWF variations, and are meant to provide a qualitative comparison with the behavior observed in the $P_{5c}$ sector.

A markedly different pattern emerges with respect to the all-charm case. 
In particular, focusing on the dominant constituent-heavy-quark channel, the $[b \to P_{5b}]$ FF exhibits a pronounced separation between the direct and scalar-diquark modes, with the direct configuration leading to significantly larger values across the explored $\mu_F$ range. 
This behavior contrasts with the $[c \to P_{5c}]$ channel, where the two modes remain closer and display a more balanced hierarchy. 
At the same time, the overall structure of subleading partonic channels departs from the pattern observed in the all-charm case. 
By contrast, the scalar-diquark mode in the bottom sector shows a more regular and suppressed behavior, closer to expectations based on heavy-mass scaling.

These features point to an enhanced sensitivity of the fully bottom system to the details of the nonperturbative modeling. 
While the formal construction of the {\tt PQ5Q1.1} framework can be extended to all-heavy configurations, the present implementation does not yet provide a sufficiently stable and robust phenomenological description in the bottom sector. 
In particular, the behavior observed in the direct $b \to P_{5b}$ channel suggests that the current compact multiquark modeling may require a dedicated reassessment when applied to fully bottom states.

For this reason, bottom-sector predictions are not included in the phenomenological analysis of this work. 
The results shown in Fig.~\ref{fig:FFs-muF_PQb} should therefore be regarded as indicative, and a systematic investigation of all-bottom pentaquark fragmentation is left for future studies.

\section{Pentacharm-jet systems within high-energy QCD}
\label{sec:hybrid_factorization}

The first part of this section (Sec.~\ref{ssec:resummation}) offers a concise overview of recent phenomenological progress in probing the semihard regime of QCD.
The second part (Sec.~\ref{ssec:NLL_cross_section}) then presents the construction of the $\NLLp$ hybrid factorization and its application to semi-inclusive $\PQc$ plus jet hadroproduction.

\subsection{Highlights of semihard phenomenology}
\label{ssec:resummation}

Heavy-flavor hadron production represents a powerful probe of QCD dynamics at high energies. In this regime, large energy logarithms spoil the convergence of fixed-order perturbative expansions, requiring an all-order resummation. The Balitsky-Fadin-Kuraev-Lipatov (BFKL) framework~\cite{Fadin:1975cb,Balitsky:1978ic} provides a systematic treatment of these effects, resumming contributions at both LL and NLL accuracy, associated with terms of the form $(\alpha_s \ln (s))^n$ and $\alpha_s (\alpha_s^n \ln (s))^n$.

Within this formalism, hadronic cross sections are written as transverse-momentum convolutions of a universal Green’s function, known at NLO~\cite{Fadin:1998py,Ciafaloni:1998gs}, with two process-dependent forward impact factors. These objects incorporate collinear ingredients such as parton distribution functions (PDFs) and FFs, leading to a \emph{hybrid} factorization scheme that consistently embeds collinear dynamics into the high-energy resummation.

The phenomenological relevance of BFKL resummation has been extensively explored, often at $\NLLp$ accuracy, in a variety of processes, including Mueller-Navelet jets~\cite{Ducloue:2013hia,Celiberto:2015yba,Celiberto:2016ygs,Celiberto:2017ius,Celiberto:2022gji,Caporale:2018qnm,Baldenegro:2024ndr}, dihadron and hadron-jet correlations~\cite{Celiberto:2016hae,Celiberto:2017ptm,Celiberto:2020rxb,Celiberto:2022rfj,Bolognino:2018oth,Celiberto:2020wpk,Celiberto:2022kxx}, multijet tags~\cite{Caporale:2015vya,Caporale:2015int,Caporale:2016soq,Caporale:2016zkc,GordoGomez:2018yjf}, forward Higgs production~\cite{Celiberto:2020tmb,Celiberto:2023rtu,Celiberto:2023uuk,Celiberto:2023eba,Celiberto:2023nym,Celiberto:2022zdg,Mohammed:2022gbk,Celiberto:2022fgx,Celiberto:2024bbv}, Drell-Yan processes~\cite{Celiberto:2018muu,Golec-Biernat:2018kem}, and heavy-hadron production~\cite{Celiberto:2021dzy,Celiberto:2021fdp,Celiberto:2022dyf,Celiberto:2023fzz,Celiberto:2022grc,Celiberto:2022keu,Celiberto:2024omj,Celiberto:2025ogy,Celiberto:2025euy,Celiberto:2026qiz,Bolognino:2021mrc}.
In addition, single-forward emissions probe the small-$x$ gluon content of the proton via the unintegrated gluon distribution (UGD), whose behavior is governed by linear BFKL evolution. Phenomenological studies of UGDs have been carried out through exclusive vector-meson production at HERA~\cite{Bolognino:2018rhb,Bolognino:2019bko} and the EIC~\cite{Bolognino:2021niq,Bolognino:2021bjd}.

The inclusion of UGD information has also improved collinear approaches, supporting the construction of small-$x$ resummed collinear~\cite{Ball:2017otu,Abdolmaleki:2018jln,Bonvini:2019wxf,Bonvini:2026cxp,Celiberto:2025nnq,Silvetti:2022hyc,Silvetti:2023suu,Rinaudo:2024hdb} and transverse-momentum-dependent~\cite{Bacchetta:2020vty,Bacchetta:2024fci,Celiberto:2019slj} PDFs.

Observables involving heavy-flavored hadrons, such as $\Lambda_c$ baryons~\cite{Celiberto:2021dzy} and singly bottomed hadrons~\cite{Celiberto:2021fdp}, have proven particularly suitable to investigate semihard dynamics at natural scales. In contrast to light-hadron production, which is affected by large NLL corrections and threshold logarithms~\cite{Bolognino:2018oth,Celiberto:2020wpk}, heavy-hadron emissions exhibit a \emph{natural stabilization} pattern~\cite{Celiberto:2022grc}, reflecting the dominant role of collinear fragmentation at high $p_T$.

This stabilization has motivated the construction of DGLAP-evolving FFs based on NRQCD inputs~\cite{Braaten:1993mp,Zheng:2019dfk,Braaten:1993rw,Zheng:2019gnb}, first for vector quarkonia~\cite{Celiberto:2022dyf,Celiberto:2023fzz} and later for charmed $B$ mesons~\cite{Celiberto:2022keu,Celiberto:2024omj}. These developments have further enabled connections with exotic hadron production~\cite{Celiberto:2023rzw,Celiberto:2024mrq,Celiberto:2024mab,Celiberto:2024beg,Gatto:2025kfl}.

\begin{figure*}[!t]
\centering
\includegraphics[width=0.75\textwidth]{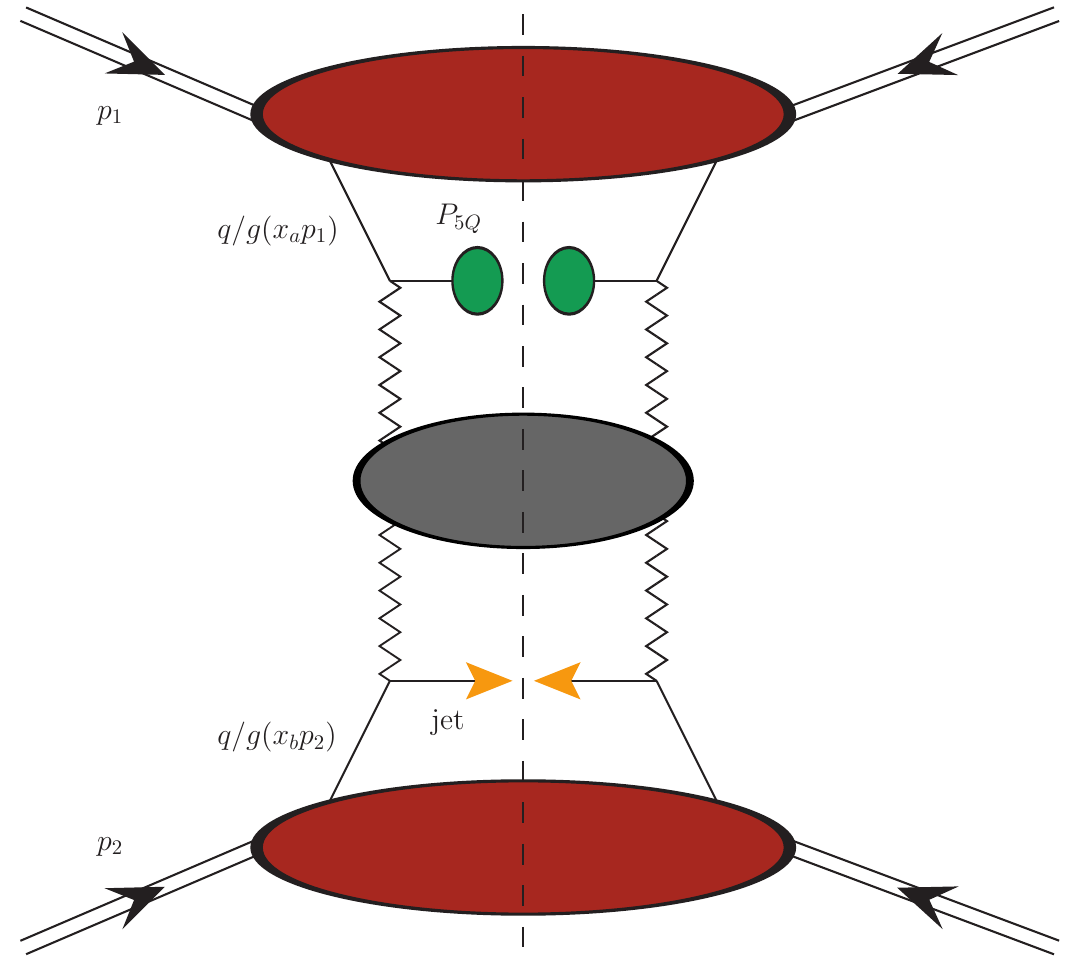}

\caption{Schematic representation of the hybrid collinear and high-energy factorization for the semi-inclusive detection of a $\PQQ$ pentaquark accompanied by a jet at hadron colliders. 
The firebrick-colored blobs represent the proton collinear PDFs, while the green ovals signify the heavy-pentaquark collinear FFs. 
Jets are indicated by orange arrows. 
The central gray blob, representing the BFKL Green’s function, is linked to the impact factors through Reggeon lines. 
All diagrams were created using {\tt JaxoDraw~2.0}~\cite{Binosi:2008ig}.}
\label{fig:reaction}
\end{figure*}

\subsection{NLL-resummed cross section}
\label{ssec:NLL_cross_section}

We examine the hadroproduction process illustrated in Fig.~\ref{fig:reaction},
\begin{equation}
\label{process}
    {\rm p}(p_1) + {\rm p}(p_2) \, \rightarrow \, \PQQ(\kappa_1, y_1, \varphi_1) + {\cal X} + {\rm jet}(\kappa_2, y_2, \varphi_2) \; ,
\end{equation}
where incoming protons with momenta $p_{1,2}$ initiate the collision. 
In this context, $\PQQ \equiv \PQc$ represents the $| c \bar{c} c c c \rangle$ pentacharm state, emitted with momentum $\kappa_1$, rapidity $y_1$, and azimuthal angle $\varphi_1$. 
Simultaneously, a light-flavor jet is detected with momentum $\kappa_2$, rapidity $y_2$, and azimuthal angle $\varphi_2$. 
The term ${\cal X}$ inclusively accounts for all additional undetected products. 
Our focus lies on the semihard and diffractive regime, characterized by large transverse momenta, $|\vec \kappa_{1,2}|$, and significant rapidity separations, $\DY = y_1 - y_2$. 
These high transverse momenta are crucial to ensuring that the ZM-VFNS treatment of leading-power fragmentation remains the primary mechanism for the formation of the $\PQQ$ states.
The parent protons' momenta establish a Sudakov basis with $p_1^2= p_2^2=0$ and $(p_1 \cdot p_2) = s/2$, leading to the relation
\begin{equation}\label{sudakov}
\kappa_{1,2} = x_{1,2} p_{1,2} - \frac{ \kappa_{1,2\perp}^2}{x_{1,2} s}p_{2,1} + \kappa_{1,2\perp} \ , \quad
\vec \kappa_{1,2}^{\,2} \equiv -\kappa_{1,2\perp}^2
\;.
\end{equation}
The connection between the longitudinal fractions ($x_{1,2}$) and the rapidities ($y_{1,2}$) of the observed particles is given by
\begin{equation}\label{y-vs-x}
y_{1,2}=\pm\frac{1}{2}\ln\frac{x_{1,2}^2 s}
{\vec \kappa_{1,2}^2 }
\qquad \mbox{and} \qquad
\drv y_{1,2} = \pm \frac{\drv x_{1,2}}{x_{1,2}}
\;.
\end{equation}
Furthermore, the following holds:
\begin{equation}
\label{DeltaY}
 \DY = y_1 - y_2 = \ln \left( \frac{x_1 x_2}{|\vec \kappa_1||\vec \kappa_2|} s \right) \;.
\end{equation}

Under a standard LO QCD collinear factorization scheme, we express the differential cross section as a collinear convolution of the on shell hard subprocess, the proton PDFs, and the pentaquark FFs
\begin{equation}
\label{sigma_collinear}
\begin{split}
\hspace{-0.25cm}
\frac{\drv\sigma^{\rm LO}_{\rm [coll.]}}{\drv x_1\drv x_2\drv ^2\vec \kappa_1\drv ^2\vec \kappa_2}
= \hspace{-0.25cm} \sum_{\alpha,\beta=q,{\bar q},g}\int_0^1 \hspace{-0.20cm} \drv x_a \int_0^1 \hspace{-0.20cm} \drv x_b\ f_\alpha\left(x_a\right) f_\beta\left(x_b\right)
\int_{x_1}^1 \hspace{-0.15cm} \frac{\drv \xi}{\xi}D^{\PQQ}_\alpha\left(\frac{x_1}{\xi}\right) 
\frac{\drv {\hat\sigma}_{\alpha,\beta}\left(\hat s\right)}
{\drv x_1\drv x_2\drv ^2\vec \kappa_1\drv ^2\vec \kappa_2}\;,
\end{split}
\end{equation}
where the summation over indices $\alpha, \beta$ includes all parton species except for the top quark, which does not contribute to fragmentation.
Here, for simplicity of notation, the explicit dependence on $\mu_F$ has been omitted from Eq.~\eqref{sigma_collinear}. 
In this expression, $f_{\alpha,\beta}\left(x_{a,b}, \mu_F \right)$ denote the proton's PDFs, while $D^{\PQQ}_\alpha\left(x_1/\xi, \mu_F \right)$ describe the parton-to-pentaquark fragmentation process. 
The longitudinal momentum fractions of the incoming partons are denoted by $x_{a,b}$, while $\xi$ represents the fraction carried by the outgoing parton producing the exotic hadron. 
Lastly, $\drv\hat\sigma_{\alpha,\beta}\left(\hat s \right)$ identifies the partonic collinear hard factors, where $\hat s \equiv x_a x_b s$ is the squared center-of-mass energy of the subprocess. 

In contrast, the high-energy resummed differential cross section within the hybrid factorization framework is described by a transverse-momentum convolution of the BFKL Green's function and two impact factors. 
We can further decompose the differential cross section into a Fourier series expansion based on the azimuthal angle difference, $\varphi = \varphi_1 - \varphi_2 - \pi$, as
\begin{equation}
 \label{dsigma_Fourier}
 \frac{\drv \sigma}{\drv y_1 \drv y_2 \drv \vec \kappa_1 \drv \vec \kappa_2 \drv \varphi_1 \drv \varphi_2} =
 \frac{1}{(2\pi)^2} \left[{\cal C}_0 + 2 \sum_{m=1}^\infty \cos (m \varphi)\,
 {\cal C}_m \right]\, .
\end{equation}
The NLL Green's function serves as the first essential element:
\begin{equation}
\label{G_BFKL_NLL}
 {\cal G}_{\rm NLL}(\DY,m,\nu,\mu_R) = e^{{\DY} \bar \alpha_s(\mu_R) \,
 \chi^{\rm NLO}(m,\nu)} \; ,
\end{equation}
with $\bar \alpha_s(\mu_R) \equiv \alpha_s(\mu_R) N_c/\pi$ and $\beta_0 = 11N_c/3 - 2 n_f/3$ being the leading coefficient of the QCD $\beta$-function. The BFKL kernel in Mellin space, denoted by $\chi$ in the exponent of Eq.~\eqref{G_BFKL_NLL}, accounts for the resummation of energy logarithms at NLL. It is given by
\begin{eqnarray}
 \label{chi}
 \chi^{\rm NLO}(m,\nu) = \chi(m,\nu) + \bar\alpha_s \tilde{\chi}(m,\nu) \;,
\end{eqnarray}
where the LO kernel eigenvalues $\chi(m,\nu)$ are
\begin{eqnarray}
 \label{kernel_LO}
 \chi\left(m,\nu\right) = -2\gamma_{\rm E} - 2 \, {\rm Re} \left\{ \psi\left(\frac{1}{2} + \frac{m}{2} + i \nu \right) \right\} \,,
\end{eqnarray}
with $\gamma_{\rm E}$ as the Euler--Mascheroni constant and $\psi(z) \equiv \Gamma^\prime(z)/\Gamma(z)$ representing the digamma function. 
The NLO correction to the kernel, $\tilde{\chi}(m,\nu)$ in Eq.\eref{chi}, is expressed as
\begin{equation}
\label{chi_NLO}
\tilde{\chi} \left(m,\nu\right) = \bar\chi(m,\nu)+\frac{\beta_0}{2 N_c}\chi(m,\nu)
\left( - \frac{1}{4}\chi(l,\nu) + \frac{5}{6} + \ln\frac{\mu_R}{\sqrt{|\vec \kappa_1||\vec \kappa_2|}} \right) \;,
\end{equation}
where $\bar\chi(m,\nu)$ was computed in Ref.~\cite{Kotikov:2000pm}.

Another fundamental component for the resummed differential cross section is the pentaquark NLO emission function, projected onto the LO kernel eigenfunctions. This follows the formulation from Ref.~\cite{Ivanov:2012iv}, which is particularly effective for heavy hadrons with high transverse momentum:
\begin{equation}
\label{PIF}
\E_{\PQQ}^{\rm NLO}(m,\nu,|\vec \kappa|,x) =
\E_{\PQQ}(\nu,|\vec \kappa|,x) +
\alpha_s(\mu_R) \, \hat \E_{\PQQ}(m,\nu,|\vec \kappa|,x) \;.
\end{equation}
The corresponding LO expression is
\begin{equation}
\label{LOPIF}
\hspace{-0.30cm}
\E_{\PQQ}(\nu,|\vec \kappa|,x) 
= 2 \sqrt{\frac{C_F}{C_A}}
|\vec \kappa|^{2i\nu-1}
\!\!\!\int_{x}^1\frac{\drv \xi}{\xi}
\left( \frac{\xi}{x} \right)
^{2 i\nu-1} 
 \!\left[\frac{C_A}{C_F}f_g(\xi)D_g^{\PQQ}\left(\frac{x}{\xi}\right)
 +\!\!\!\sum_{\alpha=q,\bar q}\!f_\alpha(\xi)D_\alpha^{\PQQ}\left(\frac{x}{\xi}\right)\right] \;,
\end{equation}
with the NLO correction, $\hat \E_{\PQQ}(m,\nu,|\vec \kappa|,x)$, found in Ref.~\cite{Ivanov:2012iv}.

Finally, we introduce the light-flavored jet impact factor:
\begin{equation}
\label{JIF}
\E_J^{\rm NLO}(m,\nu,|\vec \kappa|,x) =
\E_J(\nu,|\vec \kappa|,x) +
\alpha_s(\mu_R) \, \hat \E_J(m,\nu,|\vec \kappa|,x) \;,
\end{equation}
whose LO limit reads as
\begin{equation}
 \label{LOJIF}
 \E_J(\nu,|\vec \kappa|,x) = 2 \sqrt{\frac{C_F}{C_A}}
 |\vec \kappa|^{2i\nu-1}\left[\frac{C_A}{C_F}f_g(x)
 +\sum_{\beta=q,\bar q}f_\beta(x)\right] \;.
\end{equation}
The NLO correction $\hat \E_J(m,\nu,|\vec \kappa|,x)$ depends on the jet algorithm. We adopt the strategy proposed in Ref.~\cite{Ivanov:2012ms} and suitably optimized for numerical analyses. 
A jet selection function is derived using the so-called small-cone approximation for cone-type algorithms (see Ref.~\cite{Colferai:2015zfa} for technical details). 
In line with recent CMS forward-jet experimental studies~\cite{Khachatryan:2016udy}, we employ a jet-cone radius of ${\cal R}_J = 0.5$.

By integrating these components, we establish the master equation for the $\NLLp$ azimuthal coefficients in the $\MSb$ renormalization scheme. 
One has

\begin{eqnarray}
\label{Cm_NLLp_MSb}
 \CmNLLp \!\! &=& \!\! 
 \frac{e^{\DY}}{s} 
 \int_{-\infty}^{+\infty} \drv \nu \, 
 {\cal G}_{\rm NLL}(\DY,m,\nu,\mu_R) \,
 \alpha_s^2(\mu_R) 
 \\ \nonumber
 \!\! &\times& \!\! \biggl\{\E_{\PQQ}^{\rm NLO}(m,\nu,|\vec \kappa_1|, x_1) \,
 [\E_J^{\rm NLO}(m,\nu,|\vec \kappa_2|,x_2)]^*
 \\ \nonumber
 \!\! &+& \!\!
 \left.
 \alpha_s^2(\mu_R) \DY \frac{\beta_0}{4 \pi} \,
 \chi(m,\nu)
 \left[\ln\left(|\vec \kappa_1| |\vec \kappa_2|\right) + \frac{i}{2} \, \frac{\drv}{\drv \nu} \ln\frac{\E_{\PQQ}}{\E_J^*}\right]
 \right\}
 \;.
\end{eqnarray}
The $\NLLp$ label denotes full NLL resummation of energy logarithms while maintaining NLO accuracy. The ``$+$'' superscript highlights the inclusion of contributions beyond the NLL level from cross terms involving NLO corrections to emission functions.
In our counting, $\NLLpp$ accuracy describes the precision of the universal BFKL kernel in resumming high-energy logarithms: $\LL$ resums $[\alpha_s \ln(s)]^n$ terms via the LO kernel, while $\NLLpp$ includes $\alpha_s [\alpha_s \ln(s)]^n$ contributions from the NLO kernel. 
This kernel is convoluted with impact factors computed at LO or NLO. 
This kernel-based convention offers a universal prescription for logarithmic counting across observables, avoiding ambiguities from process-dependent impact factors (for instance, the leading-$\alpha_s$ power of impact factors differs between Higgs and jet/hadron production~\cite{Celiberto:2020tmb}).
For comparative purposes, we evaluate the pure LL limit in the $\MSb$ scheme as
\begin{equation}
\label{Cm_LL_MSb}
 \CmLL = 
 \frac{e^{\DY}}{s} 
 \int_{-\infty}^{+\infty} \drv \nu \, 
 e^{{\cal G}_{\rm NLL}^{(0)}(\DY,m,\nu,\mu_R)} 
 \alpha_s^2(\mu_R) \, \E_{\PQQ}(m,\nu,|\vec \kappa_1|, x_1)[\E_J(m,\nu,|\vec \kappa_2|,x_2)]^* \;.
\end{equation}

To compare high-energy resummation with DGLAP predictions, observables must be evaluated using both hybrid factorization and fixed-order computations. 
Given the current absence of numerical codes for fixed-order distributions in inclusive hadron-plus-jet hadroproduction at NLO accuracy, we use a high-energy fixed-order approach to assess the resummation's impact. 
This method, originally developed for dijet~\cite{Celiberto:2015yba} and hadron-jet~\cite{Celiberto:2020wpk} azimuthal correlations, truncates the high-energy series at NLO accuracy, reproducing the high-energy behavior of a pure NLO calculation. 
We limit the expansion of azimuthal coefficients in Eq.~(\ref{Cm_NLLp_MSb}) to ${\cal O}(\alpha_s^3)$, yielding the high-energy fixed-order ($\HENLOp$) expression
\begin{align}
\label{Cm_HENLOp_MSb}
 \CmHENLOp &= 
 \frac{e^{\DY}}{s} 
 \int_{-\infty}^{+\infty} \drv \nu \, 
 \alpha_s^2(\mu_R) \,
 \left[ 1 + {\cal G}_{\rm NLL}^{(0)}(\DY,m,\nu,\mu_R) \right]
 \\ \nonumber
 &\times
 \E_{\PQQ}^{\rm NLO}(m,\nu,|\vec \kappa_1|, x_1)[\E_J^{\rm NLO}(m,\nu,|\vec \kappa_2|,x_2)]^* \;,
\end{align}
where
\begin{equation}
\label{G_BFKL_0}
 {\cal G}_{\rm NLL}^{(0)}(\DY,m,\nu,\mu_R) = \bar \alpha_s(\mu_R) \DY \chi(m,\nu)
\end{equation}
represents the first-order expansion of the BFKL kernel.

Factorization ($\mu_F$) and renormalization ($\mu_R$) scales are chosen based on the natural energy scales of the final state. 
We set $\mu_F = \mu_R = \mu_N$, with $\mu_N = m_{\PQQ \perp} + |\vec \kappa_2|$ as the natural reference scale. 
Here, $m_{\PQQ \perp} = \sqrt{m_{\PQQ}^2 + |\vec \kappa_1|^2}$ is the pentaquark transverse mass. 
This single-scale scheme, summing $m_{1 \perp}$ and $m_{2 \perp}$, is a common standard in precision QCD numerical codes (\emph{e.g.}, Refs.~\cite{Alioli:2010xd,Campbell:2012am}). 
MHOUs are estimated by varying $\mu_F$ and $\mu_R$ between $\mu_N/2$ and $2\mu_N$ via the $C_\mu$ parameter.

\section{Hadron-collider phenomenology}
\label{sec:phenomenology}

Numerical results presented in this work are obtained using the \textsc{Python}+\textsc{Fortran} {\Jethad} multimodular interface~\cite{Celiberto:2020wpk,Celiberto:2022rfj,Celiberto:2023fzz,Celiberto:2024mrq,Celiberto:2024swu,Celiberto:2025csa}.
For proton PDFs, we employ the {\tt NNPDF4.0} NLO set~\cite{NNPDF:2021uiq}, accessed via {\tt LHAPDF6}~\cite{Buckley:2014ana}.

\subsection{Selection of observables: Kinematic cuts and uncertainty estimates}
\label{ssec:obserbables}

Our phenomenological analysis focuses on a single observable, namely the rapidity-interval distribution, defined as the cross section differential in the rapidity separation $\DY = y_1 - y_2$ between the produced pentaquark and the associated light jet.

This observable is directly related to the azimuthal-angle coefficient ${\cal C}_0$, introduced in Sec.~\ref{ssec:NLL_cross_section}, and integrated over the final-state phase space at fixed $\DY$. 
It reads ad
\\
\begin{equation}
 \label{obs:I}
 \frac{\drv \sigma(\DY, s)}{\drv \DY} =
 \int_{|\vec \kappa_1|^{\rm min}}^{|\vec \kappa_1|^{\rm max}} 
 \!\!\drv |\vec \kappa_1|
 \int_{|\vec \kappa_2|^{\rm min}}^{|\vec \kappa_2|^{\rm max}} 
 \!\!\drv |\vec \kappa_2|
 \int_{\max \, (y_1^{\rm min}, \, \DY + y_2^{\rm min})}^{\min \, (y_1^{\rm max}, \, \DY + y_2^{\rm max})} \drv y_1
 \, \,
 {\cal C}_0^{\rm [res]}
\Bigm \lvert_{y_2 \;=\; y_1 - \DY}
 \;,
\end{equation}
\\
where the label ${\rm [res]}$ collectively denotes $\NLLp$, $\HENLOp$, or $\LL$ accuracy. 
The integration over one rapidity variable is removed by imposing the kinematic constraint $\DY = y_1 - y_2$.

The transverse momenta of the forward pentacharm are selected in the range $50 < |\vec \kappa_1|/{\rm GeV} < 120$, ensuring the applicability of a ZM-VFNS fragmentation framework, where all relevant scales lie well above the charm evolution threshold. 
The associated jet is tagged within $30 < |\vec \kappa_2|/{\rm GeV} < 120$, in line with current LHC and HL-LHC analyses~\cite{Khachatryan:2016udy}.

Adopting asymmetric transverse-momentum cuts plays a crucial role in enhancing the sensitivity to high-energy dynamics. 
In particular, this configuration reduces the impact of fixed-order contributions~\cite{Celiberto:2015yba,Celiberto:2020wpk}, suppresses Sudakov logarithms associated with nearly back-to-back emissions~\cite{Mueller:2013wwa,Marzani:2015oyb}, and stabilizes the perturbative behavior of the cross section~\cite{Andersen:2001kta}, while mitigating violations of energy-momentum conservation~\cite{Ducloue:2014koa}.

Regarding rapidity acceptance, we adopt representative LHC-inspired cuts. 
Pentaquarks are assumed to be detected within the central region, $-2.5 < y_1 < 2.5$, corresponding to the barrel calorimeter coverage~\cite{Chatrchyan:2012xg}, while jets are allowed to extend into the forward region, $-4.7 < y_2 < 4.7$, consistent with current experimental capabilities~\cite{Khachatryan:2016udy}.

A robust phenomenological analysis demands a controlled assessment of theoretical uncertainties.  
In this work, we adopt a structured approach in which the dominant sources of uncertainty are systematically identified and treated, allowing us to disentangle their individual impact as well as to evaluate their combined effect on collider-level observables.  
The main contributions considered are the following:

\begin{itemize}

 \item[$(i)$]
 \textbf{Perturbative H-MHOUs}. 
 These originate from the residual dependence on the renormalization and factorization scales entering the hard-scattering subprocess.  
 Their effect is estimated by varying the corresponding scales around their central values by a factor between $1/2$ and $2$, thereby probing the size of missing higher-order contributions in the high-energy resummed cross section.

 \item[$(ii)$]
 \textbf{Perturbative F-MHOUs}.
 These uncertainties are associated with the perturbative input of the fragmentation functions at their initial scale.  
 As discussed in Sec.~\ref{sec:FFs}, they are modeled through variations of the evolution scales entering the FF construction, and are encoded in the replica structure of the {\tt PQ5Q1.1} set.

 \item[$(iii)$]
 \textbf{Nonperturbative F-NPWF uncertainties}.  
 These reflect the modeling of the nonperturbative wave function entering the initial-scale FFs.  
 In the present framework, they are controlled by variations of the transverse-momentum parameter $\vqTTa$, which governs the effective momentum distribution of the bound state within the Suzuki-inspired approach.  

 \item[$(iv)$]
 \textbf{Proton PDFs}. 
 An additional source of uncertainty arises from the proton PDFs, which are extracted from global fits and therefore intrinsically nonperturbative.  
 However, dedicated studies in analogous tetraquark-production channels have shown that variations across different PDF sets or replicas remain at the percentage level.  
 For this reason, and given the dominance of fragmentation-related uncertainties, we restrict our analysis to the central member of the {\tt NNPDF4.0} set~\cite{NNPDF:2021uiq}.

 \item[$(v)$]
 \textbf{Phase-space numerical integration}.
 The leading numerical uncertainty originates from the multidimensional integration over the final-state phase space (see Eq.~\eqref{obs:I}) and over the Mellin variable $\nu$ (see Eqs.~\eqref{Cm_NLLp_MSb}, \eqref{Cm_LL_MSb}, and~\eqref{Cm_HENLOp_MSb}).  
 These integrals are performed using the native routines of {\Jethad}, with numerical errors systematically kept below the $1\%$ level.  
 Subleading contributions arise from the integration over the longitudinal momentum fractions entering the PDF--FF convolution (see Eq.~\eqref{LOPIF}), and are found to be negligible.

\end{itemize}

In practice, F-MHOU and F-NPWF effects are simultaneously encoded within the replica ensemble, which provides a unified framework for uncertainty propagation (see Sec.~\ref{sec:FFs} and Appendix~\hyperlink{app:B}{B}).  
At the level of rapidity distributions, these contributions are consistently combined with H-MHOUs, allowing for a comprehensive assessment of the total theoretical uncertainty affecting the observable.

\subsection{Precision study: Numerical findings and discussion}
\label{ssec:results}

\begin{figure*}[!t]
\centering

   \hspace{0.00cm}
   \includegraphics[scale=0.395,clip]{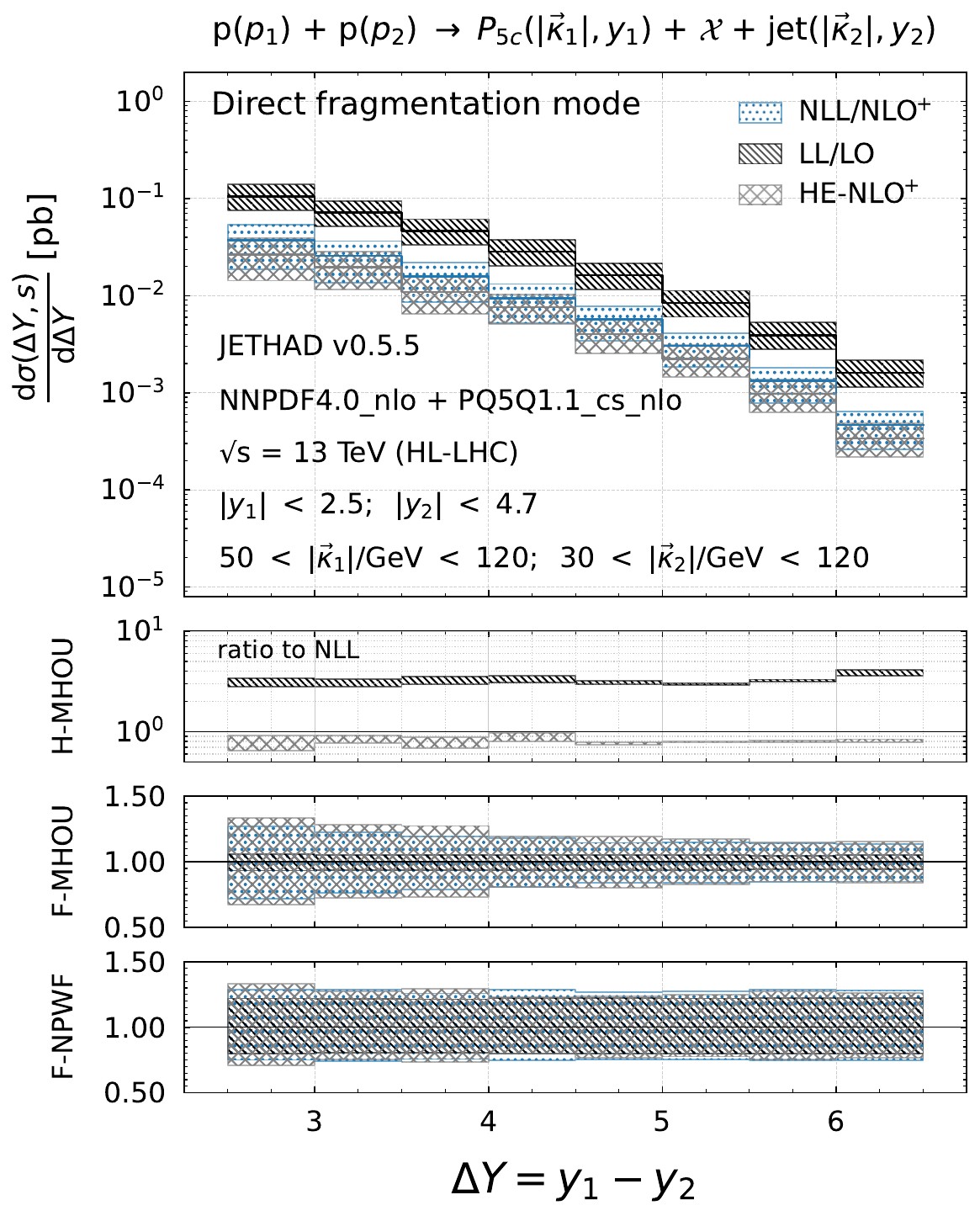}
   \hspace{0.30cm}
   \includegraphics[scale=0.395,clip]{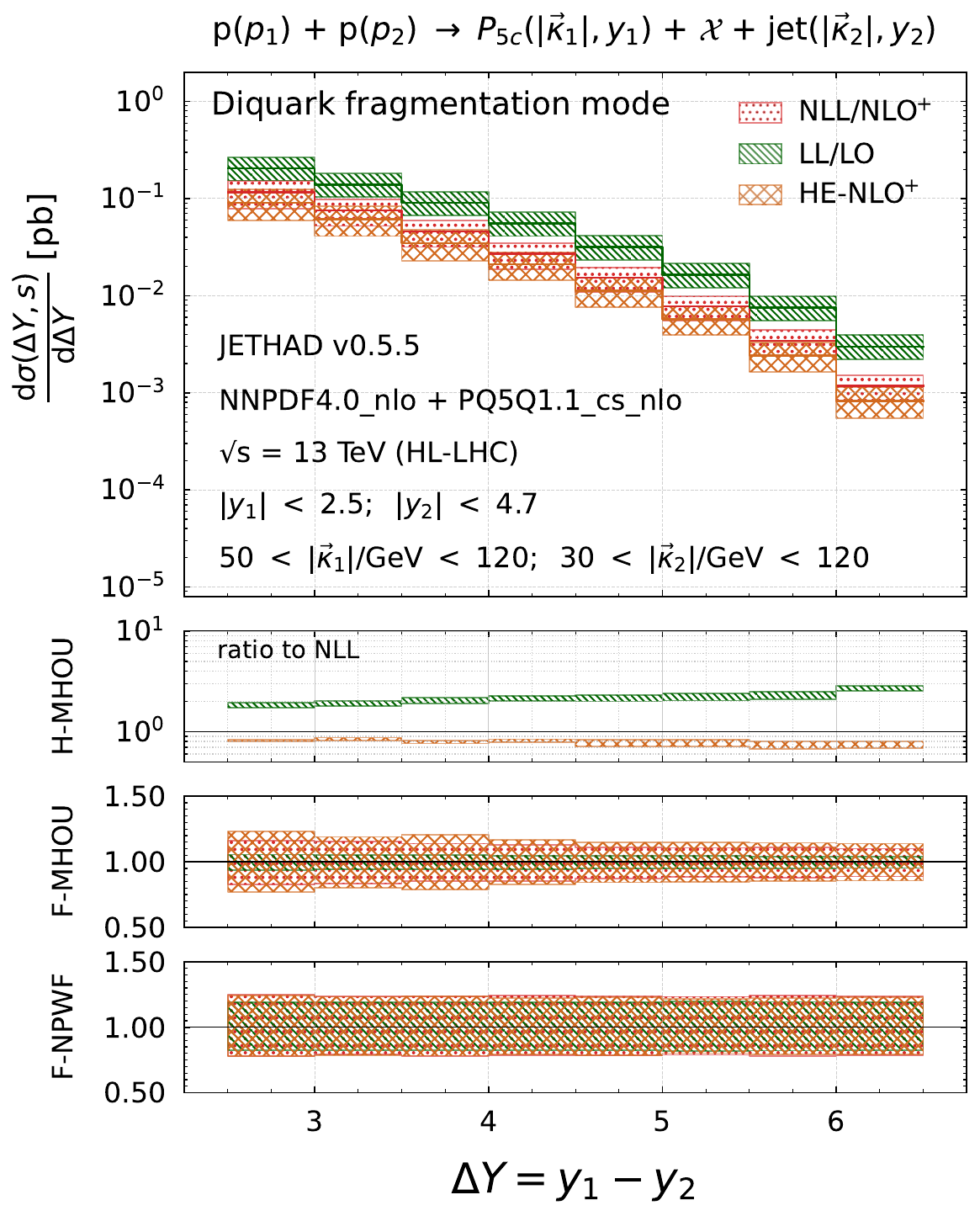}

\caption{$\DY$ distributions for semi-inclusive $\PQc$ plus jet production at $\sqrt{s} = 13$~TeV (HL-LHC), in the direct fragmentation mode (left) and in the scalar-diquark mode (right). 
Shaded bands in the main panels represent the total uncertainty, obtained by combining H-MHOU, F-MHOU, F-NPWF, and phase-space integration effects. 
The ancillary panels display: $(i)$ the ratios of $\LL$ and $\HENLOp$ predictions to the $\NLLp$ baseline, including H-MHOU variations only, and $(ii)$--$(iii)$ F-MHOU and F-NPWF uncertainties, shown as ratios to the central prediction.}
\label{fig:I_LHC}
\end{figure*}

\begin{figure*}[!t]
\centering

   \hspace{0.00cm}
   \includegraphics[scale=0.395,clip]{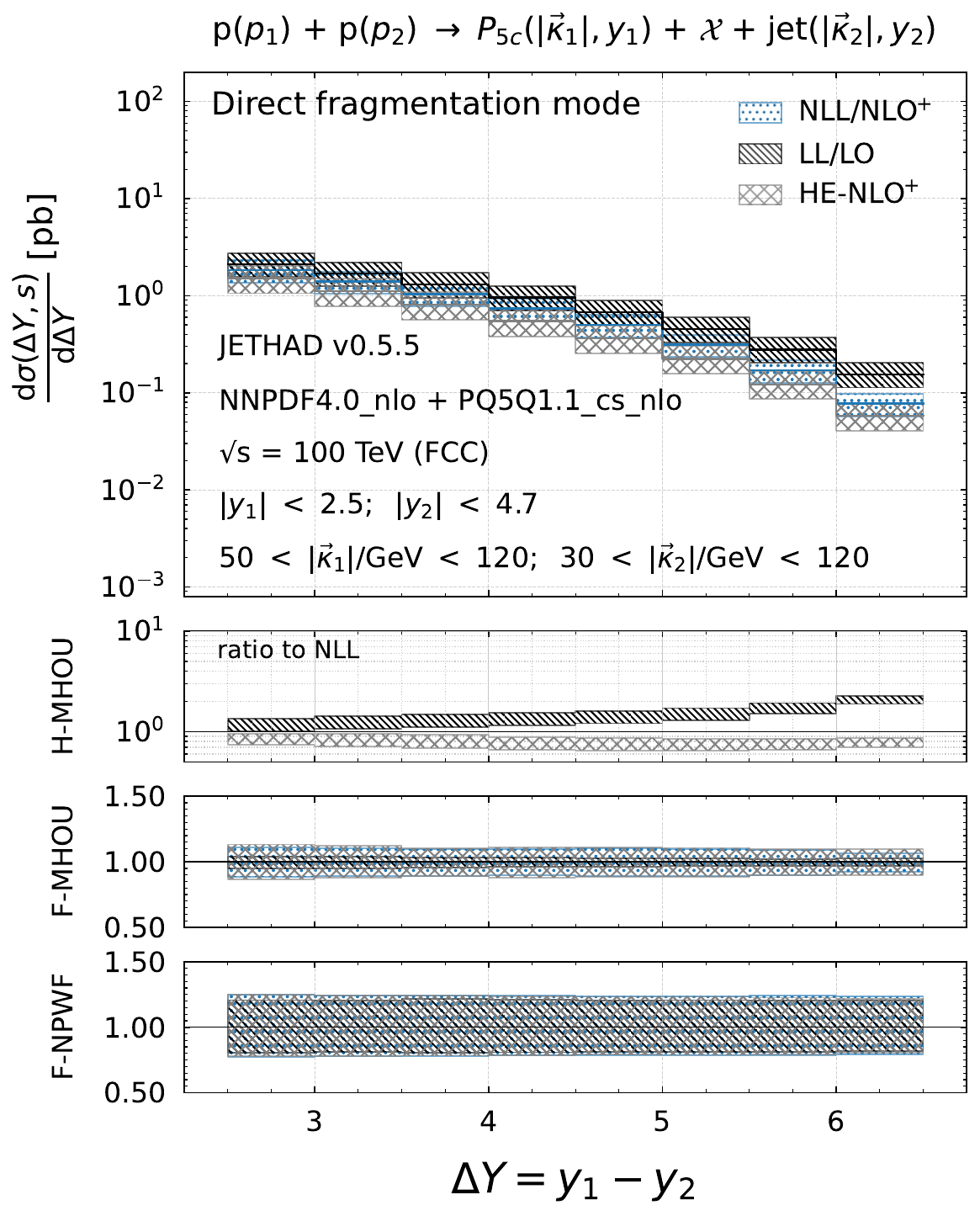}
   \hspace{0.30cm}
   \includegraphics[scale=0.395,clip]{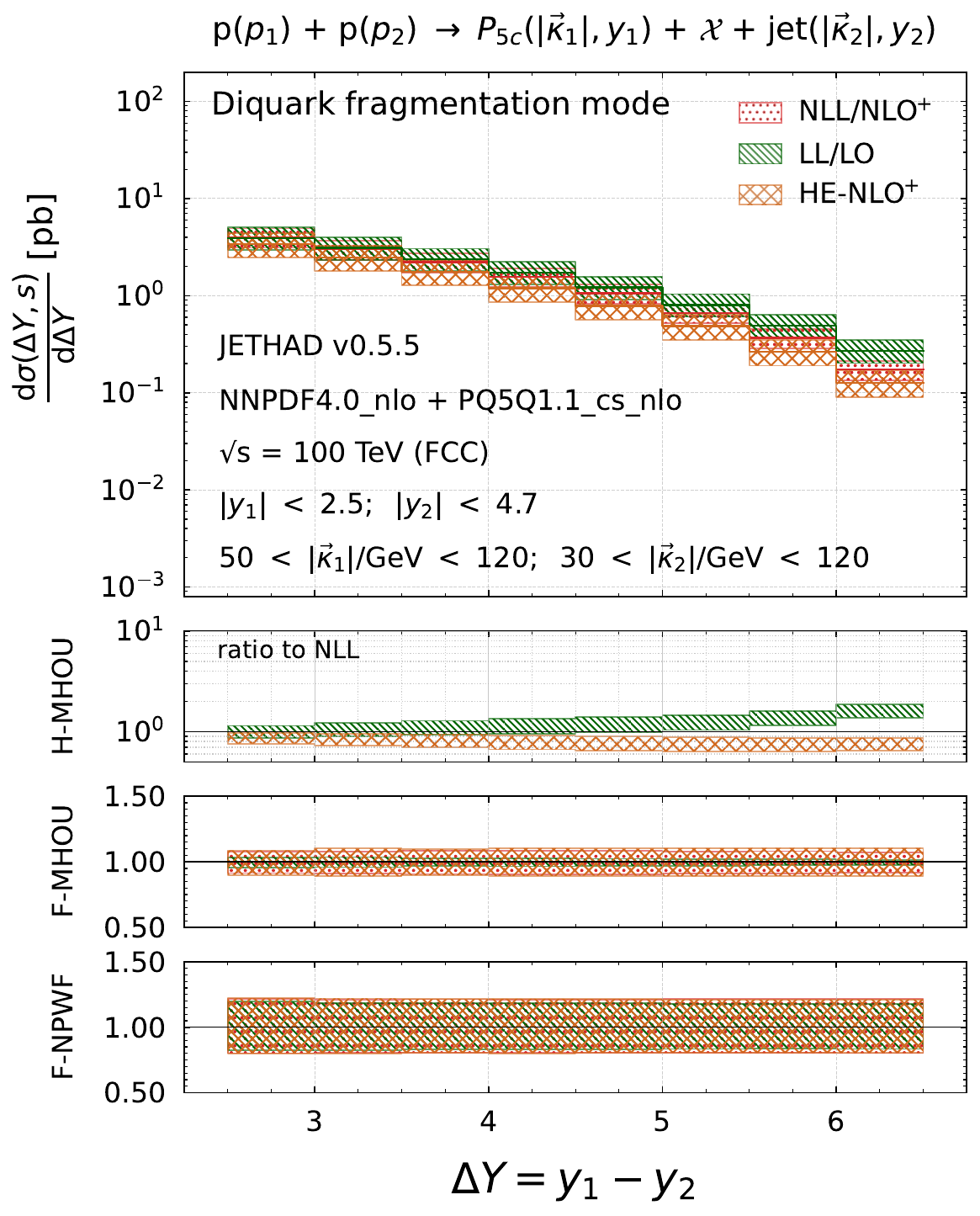}

\caption{$\DY$ distributions for semi-inclusive $\PQc$ plus jet production at $\sqrt{s} = 100$~TeV (FCC, nominal), in the direct fragmentation mode (left) and in the scalar-diquark mode (right). 
Shaded bands in the main panels represent the total uncertainty, obtained by combining H-MHOU, F-MHOU, F-NPWF, and phase-space integration effects. 
The ancillary panels display: $(i)$ the ratios of $\LL$ and $\HENLOp$ predictions to the $\NLLp$ baseline, including H-MHOU variations only, and $(ii)$--$(iii)$ F-MHOU and F-NPWF uncertainties, shown as ratios to the central prediction.}

\label{fig:I_FCC}
\end{figure*}

We present in Figs.~\ref{fig:I_LHC} and~\ref{fig:I_FCC} our predictions for the rapidity-interval distributions of semi-inclusive $\PQc$ plus light-jet production at $\sqrt{s}=13$~TeV (HL-LHC) and $\sqrt{s}=100$~TeV (FCC, nominal), respectively. 
In both cases, results are shown for the direct (left panels) and scalar-diquark (right panels) fragmentation modes.
The distributions are evaluated in uniform $\DY$ bins of width $0.5$, in order to facilitate realistic comparisons with future experimental analyses.

All distributions exhibit a monotonic decrease as $\DY$ increases. 
This behavior follows from the competition between high-energy dynamics and collinear suppression effects. 
On the one hand, the BFKL Green's function enhances the cross section as the rapidity interval grows. 
On the other hand, the convolution with proton PDFs and pentaquark FFs in the impact factors increasingly suppresses the available phase space at large $\DY$. 
In the kinematic domain considered here, the latter effect dominates, leading to a decreasing rapidity spectrum over the full range explored.

The overall normalization is significantly enhanced when moving from the HL-LHC to the FCC regime, with an increase exceeding 1 order of magnitude and reaching up to nearly 2 orders of magnitude in the low- and intermediate-$\Delta Y$ region.
At $\sqrt{s}=13$~TeV, the distributions typically span the range from about $10^{-4}$--$10^{-3}$~pb at large $\DY$ up to the $10^{-2}$--$10^{-1}$~pb level at smaller rapidity intervals. 
At $\sqrt{s}=100$~TeV, the same observable reaches the pb range in the low-$\DY$ region and remains sizable even at larger rapidity separations. 
This enhancement is consistent with the larger partonic luminosity and enlarged high-energy phase space available at FCC energies. 
Compared with the previous {\tt PQ5Q1.0} analysis~\cite{Celiberto:2025ipt}, the different transverse-momentum cuts adopted here, especially the lower jet threshold, also contribute to increasing the rates.

A clear multimodal pattern emerges from the comparison between the two fragmentation scenarios. 
The scalar-diquark mode systematically yields larger cross sections than the direct one, while preserving the same qualitative $\DY$ dependence. 
At the HL-LHC, the enhancement is particularly visible in the low- and intermediate-$\DY$ regions, where the diquark/direct ratio is typically of order of a few. 
At FCC energies, the hierarchy persists but becomes slightly more moderate and smoother across the full rapidity interval. 
This indicates that the multimodal structure of the {\tt PQ5Q1.1} FFs is not washed out by the high-energy convolution, but remains visible at the level of collider observables.

The first ancillary panels in Figs.~\ref{fig:I_LHC} and~\ref{fig:I_FCC} provide a direct assessment of the perturbative stability of the high-energy resummation. 
They show the ratios of $\LL$ and $\HENLOp$ predictions to the $\NLLp$ baseline, including H-MHOU variations only. 
At the HL-LHC, the $\LL$ result lies significantly above the $\NLLp$ one, especially in the direct mode, where the ratio can reach values of several units at intermediate and large $\DY$. 
This confirms that NLL corrections play a quantitatively important role in stabilizing the resummed prediction. 
The $\HENLOp$ curves, instead, remain much closer to the $\NLLp$ baseline, often lying below or near it, which supports the consistency of the matched high-energy expansion.

At FCC energies, the pattern becomes more stable. 
The $\LL/\NLLp$ ratio is closer to unity, particularly in the scalar-diquark mode, while the $\HENLOp/\NLLp$ ratio remains well controlled across the spectrum. 
This behavior suggests that the $\NLLp$ framework becomes increasingly natural in the higher-energy regime, where the semihard kinematics more efficiently activates the high-energy logarithmic dynamics. 
In this sense, the FCC setup offers not only larger rates, but also a cleaner environment for testing the resummed structure of the calculation.

The second and third ancillary panels isolate the two fragmentation-related sources of uncertainty introduced in the {\tt PQ5Q1.1} framework. 
F-MHOUs, associated with perturbative scale variations in the FF construction, remain comparatively small over the entire rapidity range. 
They typically induce variations at the level of a few tens of percentage or less, with a mild dependence on $\DY$. 
This indicates that the perturbative component of the fragmentation input and its subsequent evolution are under good control. 
The effect is particularly regular at FCC energies, where the F-MHOU ratios stay close to unity for both fragmentation modes.

F-NPWF uncertainties exhibit a different pattern. 
They are generally larger than F-MHOUs and display a weaker dependence on $\DY$. 
This is physically expected: F-NPWF variations mainly affect the $z$-dependence of the fragmentation functions and are propagated to the observable through the convolution, without modifying the underlying high-energy dynamics that controls the $\DY$ behavior.
Consequently, F-NPWF effects behave approximately as controlled normalization uncertainties on the rapidity distributions. 
Their impact is nevertheless moderate and stable, with no sign of pathological enhancement in the explored kinematic region.

The total uncertainty bands in the main panels combine H-MHOU, F-MHOU, F-NPWF, and phase-space integration effects. 
This decomposition is one of the main advantages of the {\tt PQ5Q1.1} framework with respect to earlier central-value analyses: it permits one to identify which theoretical ingredients control the uncertainty of the observable, rather than assigning a single global band with no internal interpretation.

From the viewpoint of high-energy phenomenology, the results confirm the natural-stability pattern already observed in heavy-flavor production. 
The dominance of heavy-quark fragmentation, together with the smooth evolution-driven gluon component discussed in Sec.~\ref{sec:FFs}, prevents the large instabilities that often affect light-particle BFKL observables. 
Although LL predictions can still differ significantly from the $\NLLp$ baseline, the matched resummed calculation remains stable, and the fragmentation-induced uncertainties are well behaved. 
This makes $\PQc$ plus jet production a promising channel for testing exotic-matter fragmentation in a high-energy environment.

Overall, Figs.~\ref{fig:I_LHC} and~\ref{fig:I_FCC} show that the {\tt PQ5Q1.1} functions lead to physically interpretable and numerically stable predictions for the all-charm pentaquark production at hadron colliders. 
The direct and scalar-diquark modes generate distinguishable but controlled phenomenological patterns, while the uncertainty decomposition provides a transparent assessment of the perturbative and nonperturbative ingredients entering the calculation. 
To our knowledge, this is the first systematic uncertainty-aware study of leading-power pentaquark fragmentation in high-energy QCD, and one of the first applications of such a framework to exotic-matter production in semi-inclusive collider observables.

\section{Final remarks}
\label{sec:conclusions}

We have developed a framework for the leading-power fragmentation of all-heavy $S$-wave pentaquark states ($\PQQ$) at present and future hadron colliders, and we introduce a new generation of \emph{multimodal} collinear fragmentation functions, the {\tt PQ5Q1.1} set.  
While the formalism is general and applicable to fully heavy systems, we focus here on the phenomenologically relevant case of all-charm pentaquarks ($\PQc$ pentacharms).  
These determinations extend and supersede the previous {\tt PQ5Q1.0} framework~\cite{Celiberto:2025ipt}, which was limited to central-value predictions, by providing for the first time an uncertainty-aware construction of pentaquark FFs.

In particular, perturbative uncertainties are quantified through missing higher-order variations, inspired by advanced scale-variation strategies developed in collinear and multidimensional PDF analyses~\cite{Kassabov:2022orn,Harland-Lang:2018bxd,Ball:2021icz,NNPDF:2024dpb}, while nonperturbative effects are probed via controlled variations of the transverse-momentum parameter entering the nonperturbative wave function.  
The resulting framework consistently embeds both sources of uncertainty within a unified replica structure.

Phenomenological predictions have been obtained through the {\Jethad} numerical interface and its symbolic extension {\psymJethad}~\cite{Celiberto:2020wpk,Celiberto:2022rfj,Celiberto:2023fzz,Celiberto:2024mrq,Celiberto:2024swu,Celiberto:2025csa}, enabling the study of semi-inclusive pentacharm-plus-jet production within the $\NLLp$ hybrid collinear and high-energy factorization scheme.  
Our analysis covers center-of-mass energies from the 13~TeV HL-LHC up to the 100~TeV FCC regime.

The adoption of a fragmentation-based description for $\PQc$ production at large transverse momentum leads to a significant stabilization of the high-energy resummation framework.  
In particular, the dominance of heavy-quark fragmentation channels mitigates instabilities associated with large NLL corrections and nonresummed threshold logarithms, giving rise to the so-called \emph{natural stability} of the perturbative expansion.  
This behavior supports the robustness and predictive power of the formalism across a wide kinematic range.

Looking ahead, further steps toward precision can be achieved by extending the $\NLLp$ framework through the inclusion of complementary resummation effects.  
In particular, connections with soft-gluon~\cite{Hatta:2020bgy} and jet-radius resummations~\cite{Dasgupta:2014yra,Banfi:2012jm,Banfi:2015pju,Liu:2017pbb}, as well as with studies of jet angularities~\cite{Caletti:2021oor}, represent promising directions for future developments.

More broadly, heavy-flavor fragmentation at leading power constitutes a privileged interface between hadronic structure and precision QCD.  
The presence of heavy quarks introduces a dual dynamical regime: while the short-distance component can be computed perturbatively, the bound-state formation requires a dedicated modeling of nonperturbative dynamics, including possible correlations among constituent partons.  
A quantitative description of heavy-hadron production thus emerges from the interplay between these two sectors.

In this perspective, the {\tt PQ5Q1.1} functions provide a versatile tool for exploratory studies of pentaquark production across a wide class of processes, ranging from semi-inclusive emissions at hadron colliders to future lepton and lepton-hadron facilities.  
At the same time, the multimodal structure of the initial-scale input offers a flexible framework to encode different production mechanisms, from \emph{compact multiquark} configurations to \emph{diquark-driven} dynamics, and potentially more extended structures.

Future extensions will include the calculation of additional partonic channels, which will allow for a fully consistent implementation of the {\HFNRevo} methodology~\cite{Celiberto:2025euy,Celiberto:2024mex,Celiberto:2024bxu,Celiberto:2024rxa,Celiberto:2025xvy,Celiberto:2026rzi,Celiberto:2026zss}, including the semianalytic decoupled evolution ({\tt EDevo}), and may provide further insight into the relative production of $\PQc$ and $\bPQc$ states.

A formal extension of the present framework to all-bottom pentaquarks is, in principle, straightforward.  
However, exploratory studies indicate that the current modeling does not yet provide a sufficiently robust phenomenological description in that sector, pointing to an increased sensitivity to the details of the nonperturbative dynamics.  
In particular, while the scalar-diquark mode exhibits a qualitatively expected suppression pattern, the direct compact configuration leads to a less natural charm-bottom hierarchy, and some FF channels develop irregular structures under evolution. 

These features may signal a nontrivial interplay between heavy-quark mass effects and the modeling of the bound-state dynamics.  
A dedicated reassessment of these aspects will be required before reliable bottom-sector predictions can be established.

A more refined understanding of hadron structure will emerge as theoretical developments are confronted with data from next-generation facilities, including the FCC~\cite{FCC:2018byv,FCC:2018evy,FCC:2018vvp,FCC:2018bvk} and other future colliders~\cite{Chapon:2020heu,Anchordoqui:2021ghd,AlexanderAryshev:2022pkx,InternationalMuonCollider:2024jyv,MuCoL:2024oxj,Black:2022cth,Accardi:2023chb}.  
In this context, recent studies have highlighted how unresolved $\Jpsi$ plus charmed-jet photoproduction at the EIC~\cite{Flore:2020jau,AbdulKhalek:2021gbh,Khalek:2022bzd,Abir:2023fpo} can provide direct sensitivity to intrinsic-charm (IC) components in the proton~\cite{Brodsky:1980pb,Ball:2016neh,Hou:2017khm,Ball:2022qks,Guzzi:2022rca,NNPDF:2023tyk}. 

In parallel, axial-vector tetraquarks have been identified as particularly sensitive probes of intrinsic charm, as their radiative production mechanism enhances the dependence on IC contributions~\cite{Celiberto:2025vra}.  
These findings suggest a natural extension toward the pentaquark sector, where similar mechanisms may establish a direct connection between intrinsic heavy-quark dynamics and the formation of multiquark states, thus bridging hadron-structure effects and exotic spectroscopy~\cite{Vogt:2024fky}.

Early investigations have already explored possible links between intrinsic charm and multiquark dynamics.  
Connections between IC effects and doubly charmed pentaquarks were discussed in Ref.~\cite{Mikhasenko:2012km}, while the role of intrinsic charm in $\Lambda_b$ decays into pentaquark states was analyzed in Ref.~\cite{Hsiao:2015nna}.  
Moreover, longstanding observations of multi-$\Jpsi$ production, starting from the NA3 experiment~\cite{NA3:1982qlq} and later supported by collider data, have been interpreted as possible hints of double intrinsic-charm mechanisms and exotic multiquark configurations.

Another promising direction concerns the interplay between exotic heavy-flavor production and the dead-cone effect in QCD~\cite{Dokshitzer:1991fd}, recently observed by the ALICE Collaboration~\cite{ALICE:2021aqk}.  
Observables involving fully heavy states, particularly in jet-inclusive environments, may provide novel access to heavy-quark radiation patterns and to the dynamics of in-medium energy loss.

In this broader perspective, the study of all-charm pentaquarks through fragmentation does not simply extend existing approaches to a new class of hadrons.  
Rather, it establishes a quantitative framework in which exotic spectroscopy, hadronic structure, and precision QCD can be simultaneously addressed.  
By embedding uncertainty-aware fragmentation into high-energy phenomenology, the {\tt PQ5Q1.1} program contributes to a paradigm shift: from exploratory modeling toward a systematically improvable description of exotic-matter production, ultimately enabling future data to constrain both the dynamics of heavy quarks and the internal structure of multiquark bound states.

\section*{Data availability}
\label{sec:data_availability}
\addcontentsline{toc}{section}{\nameref{sec:data_availability}}

The {\tt PQ5Q1.1} collinear FFs~\cite{Celiberto:2026_PQ5Q11} for all-charm $S$-wave pentaquarks are publicly available at \url{https://github.com/FGCeliberto/Collinear_FFs/}. 
They are distributed in {\tt LHAPDF} format, including the full replica set.

\section*{Acknowledgments}
\label{sec:acknowledgments}
\addcontentsline{toc}{section}{\nameref{sec:acknowledgments}}

We thank  A.~Pilloni and S.~M.~Moosavi~Nejad for valuable conversations on exotic hadron physics and pentaquark fragmentation, respectively. 
We also acknowledge the members of the Quarkonia As Tools and EXOTICO communities for many motivating discussions. 
This research has been supported by the Atracción de Talento Grant n. 2022-T1/TIC-24176, funded by the Comunidad Autónoma de Madrid, Spain.
\appendix

\counterwithin*{equation}{section}
\renewcommand\theequation{\thesection\arabic{equation}}

\counterwithin*{table}{section}
\renewcommand\thetable{\thesection\arabic{table}}

\hypertarget{app:A}{
\section{Direct fragmentation coefficients}
}
\label{app:A}

In the context of the direct multiquark scenario, the fragmentation coefficients $\gamma_{P,\,{\rm [direct]}}^{(Q)}(z; k)$ involved in Eq.~\eqref{PQQ_FF_initial-scale_Q_num_direct} are expressed by

\begin{subequations}
\allowdisplaybreaks 
\begin{align}
 \label{aA:eq:gamma_k}
 \nonumber
 \gamma_{P,\,{\rm [direct]}}^{(Q)}(z; 0) &= (z+5)^2 (347638095 z^{18}-14333286144 z^{17}+303132501680 z^{16} \\ \nonumber
\,&-\, 
 2979357459428 z^{15}+20140712527168 z^{14}-101570323071060 z^{13} \\ \nonumber
\,&+\, 
 381426354056100 z^{12}-1070325481916500 z^{11}+2274630737903750 z^{10} \\
\,&-\, 
 3716109400562500 z^9+4726702575562500 z^8-4727934482187500 z^7 \\ \nonumber
\,&+\, 
 3748290871875000 z^6-2366985523437500 z^5+1187884179687500 z^4 \\ \nonumber
\,&-\, 
 464519335937500 z^3+134047607421875 z^2 \\ \nonumber
\,&-\, 
 25307617187500 z+2319335937500) \;, \\[0.50cm] \nonumber
 \gamma_{P,\,{\rm [direct]}}^{(Q)}(z; 1) &= 1133852889 z^{18}-26979532892 z^{17}+241365645652 z^{16} \\ \nonumber
\,&-\, 
 406276817520 z^{15}-5088767528028 z^{14}+35085927681040 z^{13}
\\ \nonumber
\,&-\, 
 85928556891600 z^{12}-39913868698000 z^{11}+926214176418750 z^{10} \\
\,&-\, 
 3208940492025000 z^9+6562195688125000 z^8-9297361556250000 z^7 \\ \nonumber
\,&+\, 
 9712589920312500 z^6-7742747343750000 z^5+4790060781250000 z^4
\\ \nonumber
\,&-\, 
 2279664843750000 z^3+792703369140625 z^2 \\ \nonumber
\,&-\, 
 178178710937500 z+19165039062500 \;, \\[0.50cm]
 \nonumber
 \gamma_{P,\,{\rm [direct]}}^{(Q)}(z; 2) &= 331172883 z^{16}-5540112194 z^{15}+24041536795 z^{14}+1829547432 z^{13} \\ \nonumber
\,&-\, 
 453100415317 z^{12}+1783755355930 z^{11}-1450496653425 z^{10} \\
\,&-\, 
 9989878632500 z^9+42225730018125 z^8-87820041068750 z^7 \\ \nonumber
\,&+\, 
 120577380890625 z^6-120190321250000 z^5+90757141015625 z^4 \\ \nonumber
\,&-\, 
 51981503906250 z^3+21462470703125 z^2 \\ \nonumber
\,&-\, 
 5640039062500 z+698242187500 \;, \\[0.50cm]
 \nonumber
 \gamma_{P,\,{\rm [direct]}}^{(Q)}(z; 3) &= 161946833 z^{14}-1713576308 z^{13}+6158065036 z^{12}+2311752868 z^{11} \\ \nonumber
\,&-\, 
 64931596275 z^{10}+119000823740 z^9+171719850500 z^8 \\
\,&-\, 
 1135906421000 z^7+2497182556875 z^6-3447541987500 z^5 \\ \nonumber
\,&+\, 
 3411658187500 z^4-2484448437500 z^3+1270333984375 z^2 \\ \nonumber
\,&-\, 
 402773437500 z+58789062500 \;, \\[0.50cm]
 \nonumber
 \gamma_{P,\,{\rm [direct]}}^{(Q)}(z; 4) &= 59591417 z^{12}-431073302 z^{11}+687052153 z^{10}+1761733340 z^9 \\
\,&-\, 
 6636037130 z^8+1595946780 z^7+28907583850 z^6-80523124000 z^5 \\ \nonumber
\,&+\, 
 124845798125 z^4-129207593750 z^3+87968828125 z^2 \\ \nonumber
\,&-\, 
 35317187500 z+6289062500 \;, \\[0.50cm]
 \nonumber
 \gamma_{P,\,{\rm [direct]}}^{(Q)}(z; 5) &= 5173335 z^{10}-23475428 z^9+16082740 z^8+88328984 z^7 \\
\,&-\, 
 200793674 z^6-4722280 z^5+759004600 z^4-1640425000 z^3 \\ \nonumber
\,&+\, 
 1744896875 z^2-964862500 z+220937500 \;, \\[0.50cm]
 \gamma_{P,\,{\rm [direct]}}^{(Q)}(z; 6) &= 121839 z^8-324922 z^7-210833 z^6+1991920 z^5-2327663 z^4 \\ \nonumber
\,&-\, 
 2475170 z^3+8563925 z^2-7864500 z+2537500 \;, \\[0.50cm]
 \gamma_{P,\,{\rm [direct]}}^{(Q)}(z; 7) &= 3277 z^6-2756 z^5-17820 z^4+30188 z^3 \\ \nonumber
\,&-\, 
 17647 z^2-65740 z+36500 \;, \\[0.50cm]
 \gamma_{P,\,{\rm [direct]}}^{(Q)}(z; 8) &= 183 z^4+262 z^3-769 z^2-532 z+1180 \;, \\[0.50cm]
 \gamma_{P,\,{\rm [direct]}}^{(Q)}(z; 9) &= (z+2)^2 \;.
\end{align}
\end{subequations}

\hypertarget{app:B}{
\section{Replica structure of the {\tt PQ5Q1.1} functions}
}
\label{app:B}

The {\tt PQ5Q1.1} replica ensemble is organized into two distinct blocks, corresponding to the direct and scalar-diquark production mechanisms. 
Replicas $0$--$8$ belong to the direct configuration, while replicas $9$--$17$ correspond to the diquark configuration.
Within each block, the replica structure is constructed through a nested scan over perturbative and nonperturbative variations. 
Perturbative uncertainties are modeled via energy-scale variations (F-MHOUs), implemented through multiplicative factors applied to the characteristic scale of the process. 
For each choice of energy-scale variation, we perform a scan over the transverse-momentum parameter $\vqTTa$, controlling the F-NPWF over the range $\vqTTa = 80,\,90,\,100~\text{GeV}^2$.
This procedure results in a structured ensemble in which each replica corresponds to a specific combination of perturbative (F-MHOU) and nonperturbative (F-NPWF) effects. 
The reference configuration corresponds to the central choice of energy scale ($K_{\mu} = 1$) and $\vqTTa = 90~\text{GeV}^2$.
The explicit mapping between replica indices and parameter configurations is summarized in Table~\ref{tab:PQ5Q11_replicas}.

\begin{table}[t]
\centering
\begin{tabular}{|c|c|r|r|}
\toprule
Replica ID & Mode & $K_{\mu}$ & $\vqTTa$ [GeV$^2$] \\
\midrule
0  & direct  & 1   & 90  \\
1  & direct  & 1   & 80  \\
2  & direct  & 1   & 100 \\
3  & direct  & 0.5 & 90  \\
4  & direct  & 0.5 & 80  \\
5  & direct  & 0.5 & 100 \\
6  & direct  & 2   & 90  \\
7  & direct  & 2   & 80  \\
8  & direct  & 2   & 100 \\
9  & diquark & 1   & 90  \\
10 & diquark & 1   & 80  \\
11 & diquark & 1   & 100 \\
12 & diquark & 0.5 & 90  \\
13 & diquark & 0.5 & 80  \\
14 & diquark & 0.5 & 100 \\
15 & diquark & 2   & 90  \\
16 & diquark & 2   & 80  \\
17 & diquark & 2   & 100 \\
\bottomrule
\end{tabular}
\caption{Structure of the {\tt PQ5Q1.1} replica set. The parameter $K_{\mu}$ encodes energy-scale variations (F-MHOUs), while $\vqTTa$ controls nonperturbative wave-function (F-NPWF) effects.}
\label{tab:PQ5Q11_replicas}
\end{table}

\clearpage
\bibliographystyle{apsrev}
\bibliography{bibliography}

\end{document}